\documentclass[reprint, 10pt, amsmath, amssymb, floatfix, aps, prl,  superscriptaddress, nofootinbib, nobibnotes,twocolumn]{revtex4-1}

\pdfoutput=1
\usepackage{graphicx}
\graphicspath{{figures/}}
\usepackage[utf8]{inputenc}
\usepackage{amsmath}
\usepackage{amsfonts}
\usepackage{amssymb}
\usepackage{multirow}
\usepackage{CJKutf8}
\usepackage{color}
\usepackage[colorlinks=true, linkcolor=red, citecolor=blue]{hyperref}
\usepackage[capitalise]{cleveref}
\usepackage{acro}
\usepackage{adjustbox}
\usepackage{array}
\usepackage{natbib}
\usepackage{dblfnote}
\DFNalwaysdouble
\usepackage{slashed}
\usepackage{dcolumn}
\usepackage{hyperref}
\usepackage{geometry}
\usepackage{inputenc}

\def\({\left(}
\def\){\right)}
\def\[{\left[}
\def\]{\right]}
\def\be{\begin{eqnarray}}
\def\ee{\end{eqnarray}}
\def\ss{\small}

\DeclareAcronym{GW}{
  short = GW ,
  long = gravitational wave ,
  short-plural = s 
}
\DeclareAcronym{LIGO}{
  short = LIGO ,
  long = Laser Interferometer Gravitational-wave Observatory ,
  short-plural = 
}
\DeclareAcronym{LISA}{
  short = LISA ,
  long = Laser Interferometer Space Antenna ,
  short-plural =  
}
\DeclareAcronym{SKA}{
  short = SKA ,
  long = Square Kilometre Array ,
  short-plural =  
}  
  
\DeclareAcronym{SNR}{
	short = SNR ,
	long = signal-to-noise ratio ,
	short-plural = 
}

\DeclareAcronym{PTA}{
	short = PTA ,
	long = pulsar timing array ,
	short-plural = 
}

\DeclareAcronym{FLRW}{
  short = FLRW ,
  long = Friedmann-Lemaitre-Robertson-Walker ,
  short-plural =  
}

\DeclareAcronym{SIGW}{
	short = SIGW ,
	long = scalar induced gravitational wave ,
	short-plural =  s
}

\DeclareAcronym{PBH}{
	short = PBH ,
	long = Primordial black holes ,
	short-plural =  s
}

\DeclareAcronym{CMB}{
	short = CMB ,
	long = cosmic microwave background ,
	short-plural =  s
}
\DeclareAcronym{DM}{
	short = DM ,
	long = dark matter ,
	short-plural =  s
}

\DeclareAcronym{LSS}{
	short = LSS ,
	long = large scale structure ,
	short-plural =  s
}

\DeclareAcronym{RD}{
	short = RD ,
	long = radiation-dominated ,
	short-plural =  
}

\begin{document}

\title{Primordial black holes and third order scalar induced gravitational waves}

\author{Zhe Chang} 
\affiliation{Institute of High Energy Physics, Chinese Academy of Sciences, Beijing 100049, China}
\affiliation{University of Chinese Academy of Sciences, Beijing 100049, China}

\author{Yu-Ting Kuang} 
\affiliation{Institute of High Energy Physics, Chinese Academy of Sciences, Beijing 100049, China}
\affiliation{University of Chinese Academy of Sciences, Beijing 100049, China}

\author{Xukun Zhang}   
\affiliation{Institute of High Energy Physics, Chinese Academy of Sciences, Beijing 100049, China}
\affiliation{University of Chinese Academy of Sciences, Beijing 100049, China}

\author{Jing-Zhi Zhou} 
\email{zhoujingzhi@ihep.ac.cn}
\affiliation{Institute of High Energy Physics, Chinese Academy of Sciences, Beijing 100049, China}
\affiliation{University of Chinese Academy of Sciences, Beijing 100049, China}


\begin{abstract}
The process of \acp{PBH} formation would be inevitably accompanied by \acp{SIGW}. This strong correlation between \acp{PBH} and \acp{SIGW} signals could be a promising approach to detecting \acp{PBH} in the upcoming \ac{GW} experiments, such as \ac{LISA}. We investigate the third order \acp{SIGW} during a \ac{RD} era in the case of a monochromatic primordial power spectrum $\mathcal{P}_{\zeta}=A_{\zeta}k_*\delta\left(k-k_*\right)$. For \ac{LISA} observations, the relations between \ac{SNR} and monochromatic primordial power spectrum are studied systematically. It shows that the effects of third order \acp{SIGW} extend the cutoff frequency from $2f_*$ to $3f_*$ and  lead to about $200\%$ increase of the \ac{SNR} for  frequency band from $10^{-5}$Hz to $1.6\times 10^{-3}$Hz corresponding to \acp{PBH} with mass range $4\times 10^{-12}M_{\odot} \sim 10^{-7}M_{\odot}$. We find that there exists a critical value $A_*=1.76\times 10^{-2}$ for the amplitude of the monochromatic primordial power spectra, such that when $A_{\zeta}>A_*$, the energy density of third order \acp{SIGW} will be larger than the energy density of second order \acp{SIGW}. 
\end{abstract}

\maketitle
\acresetall

\emph{Introduction.}---The inflation theory predicts that the cosmological perturbations are originated from the quantum fluctuations during inflation, which means that valuable information about the early Universe is encoded in these perturbations. On large scales ($\gtrsim$1 Mpc),  the  primordial curvature perturbations have been determined through the observations of the \ac{CMB} and \ac{LSS} \cite{Planck:2018vyg,Abdalla:2022yfr}.  It indicates a nearly scale-invariant power spectrum of primordial curvature perturbations with amplitude $\sim 2\times 10^{-9}$. On small scales ($\lesssim$1 Mpc), the constraints of primordial curvature perturbations are significantly weaker than those on large scales \cite{Bringmann:2011ut}. 

Over the past few years, the primordial scalar perturbations with large amplitudes on small scales have been attracting a lot of interests on account of their rich and profound phenomenology. If the amplitude of the small-scale primordial curvature perturbations are large enough, large-amplitude perturbation modes that enter the Hubble radius during the \ac{RD} era can lead to the production of \acp{PBH} \cite{Sasaki:2018dmp,Carr:1975qj,Carr:1993aq,Carr:2016drx,Carr:2020gox,Carr:2020xqk,Garcia-Bellido:2017mdw,Ivanov:1994pa,DeLuca:2020agl,DeLuca:2020qqa,Vaskonen:2020lbd,Braglia:2020eai,Bartolo:2018evs,Byrnes:2018clq,Ballesteros:2017fsr,Ezquiaga:2017fvi,Inomata:2016rbd,Kawasaki:2016pql,Young:2013oia,Gow:2020bzo,Wang:2016ana,Wang:2021djr}, which is a reasonable candidate for the whole or an appreciable portion of \ac{DM}. Moreover, the process during which the \acp{PBH} are formed would be inevitably accompanied by \acp{SIGW} \cite{Mollerach:2003nq,Ananda:2006af,Baumann:2007zm,Kohri:2018awv}.

The \acp{SIGW} have been studied for many years \cite{Domenech:2021ztg}. The energy density spectra of \acp{SIGW} include valuable information about \acp{PBH} \cite{Saito:2008jc,Wang:2019kaf,Inomata:2018epa,Byrnes:2018txb,Garcia-Bellido:2017aan,Nakama:2016gzw,Bugaev:2010bb,Saito:2009jt,Bugaev:2009zh,Inomata:2020lmk,Ballesteros:2020qam,Lin:2020goi,Chen:2019xse,Cai:2019elf,Cai:2019jah,Ando:2018qdb,Di:2017ndc,Chang:2022dhh,Chang:2022vlv,Orlofsky:2016vbd,Gao:2021vxb,Nakama:2015nea,Nakama:2016enz,Changa:2022trj,Inomata:2020cck} and primordial non-Gaussianity \cite{Cai:2018dig,Atal:2021jyo,Zhang:2020uek,Yuan:2020iwf,Davies:2021loj,Rezazadeh:2021clf,Kristiano:2021urj,Bartolo:2018qqn}. Furthermore, recent studies on \acp{SIGW} were also extended to gauge issue \cite{Hwang:2017oxa,Yuan:2019fwv,Inomata:2019yww,DeLuca:2019ufz,Domenech:2020xin,Chang:2020tji,Ali:2020sfw,Lu:2020diy,Tomikawa:2019tvi,Gurian:2021rfv,Uggla:2018fiy,Domenech:2021and}, dumping effect \cite{Zhang:2022dgx,Mangilli:2008bw,Saga:2014jca}, epochs of the Universe \cite{Papanikolaou:2020qtd,Domenech:2020kqm,Domenech:2019quo,Inomata:2019zqy,Inomata:2019ivs,Assadullahi:2009nf,Witkowski:2021raz,Dalianis:2020gup,Hajkarim:2019nbx,Bernal:2019lpc,Das:2021wad,Haque:2021dha}, and modified gravity \cite{Ananda:2007xh,Papanikolaou:2021uhe,Papanikolaou:2022hkg}. The relation between the \acp{SIGW} and \acp{PBH} formation was first studied in Ref.~\cite{Saito:2008jc} in the case of a monochromatic primordial power spectrum. In those previous studies, the second order \ac{SIGW} were considered only. In Ref.~\cite{Yuan:2019udt}, the authors partly considered the third order effects of \acp{SIGW} which are induced by first order scalar perturbation directly. They found that effects of the third-order correction lead to almost $20\%$ increase of the \ac{SNR} for \ac{LISA}  observations. Their results indicate that the higher order effects are not dispensable, and it is necessary to consider the higher order \acp{SIGW} when the small-scale primordial curvature perturbations are large enough. In Ref.~\cite{Zhou:2021vcw}, we studied the third order \acp{SIGW} during the \ac{RD} era systematically. In addition to the source terms of the first order scalar perturbation, the source terms of three kinds of second order perturbations induced by the first order scalar perturbation were considered completely. We found that the third order gravitational waves sourced by the second order scalar perturbations dominate the energy density spectra of third order \acp{SIGW}. The direct contributions of the source term of the first order scalar perturbation which were studied in Ref.~\cite{Yuan:2019udt} are negligible compared to the total energy density spectrum of the third order \acp{SIGW}. 

In this paper, we investigate the primordial curvature perturbations and \acp{PBH} in terms of the \acp{SIGW}. We find that the effects of the third order \acp{SIGW} have significant observational implications for \ac{LISA}.

\emph{Third order \acp{SIGW}}---The perturbed metric in the \ac{FLRW} spacetime with
Newtonian gauge takes the form
\begin{equation}
	\begin{aligned}
		\mathrm{d} s^{2}&=-a^{2}\Bigg[\left(1+2 \phi^{(1)}+ \phi^{(2)}\right) \mathrm{d} \eta^{2}+ V_i^{(2)} \mathrm{d} \eta \mathrm{d} x^{i} \\
		&+\left(\left(1-2 \psi^{(1)}- \psi^{(2)}\right) \delta_{i j}+\frac{1}{2} h_{i j}^{(2)}+\frac{1}{6} h_{i j}^{(3)}\right)\mathrm{d} x^{i} \mathrm{d} x^{j}\Bigg] \ ,
	\end{aligned}
\end{equation}
where $\phi^{(n)}$ and $\psi^{(n)}$$(n=1,2)$ are the $n$-order scalar perturbations,  $V^{(2)}_i$ is the second order vector perturbation, and $h_{i j}^{(n)}$ $(n=2,3)$ are the $n$-order tensor perturbations. In the \ac{RD} era, the first order scalar perturbation is given by \cite{Baumann:2007zm,Kohri:2018awv}
\begin{equation}
	\psi(\eta,\mathbf{k}) = \phi(\eta,\mathbf{k})=\frac{2}{3}\zeta_{\mathbf{k}} T_\phi(k \eta) \ , 
\end{equation}
where $\zeta_{\mathbf{k}}$ is the primordial curvature perturbations. The transfer functions $ T_\phi(k \eta)$ in the RD era is 
\begin{equation}\label{eq:T}
	T_{\phi}(x)=\frac{9}{x^{2}}\left(\frac{\sqrt{3}}{x} \sin \left(\frac{x}{\sqrt{3}}\right)-\cos \left(\frac{x}{\sqrt{3}}\right)\right) \ , 
\end{equation}
where we have defined $x\equiv k\eta$. We simplify the higher order cosmological perturbations in terms of the \texttt{xPand} package \cite{Pitrou:2013hga} and obtain the equation of motion of third order \acp{SIGW}
\begin{equation}\label{eq:h3}
	h_{i j}^{(3)''}+2 \mathcal{H}  h_{i j}^{(3)'}-\Delta h_{i j}^{(3)}=-12 \Lambda_{i j}^{l m} \mathcal{S}^{(3)}_{l m} \ ,
\end{equation}
where $\Lambda_{i j}^{l m}=\mathcal{T}_{i}^{l} \mathcal{T}_{j}^{m}-\frac{1}{2} \mathcal{T}_{i j} \mathcal{T}^{l m}$ is the transverse and traceless operator \cite{Chang:2020tji,Zhou:2021vcw}, and $\mathcal{T}_{i}^{l}$ is defined as $\mathcal{T}_{i}^{l}=\delta_{i}^{l}-\partial^{l} \Delta^{-1} \partial_{i}$. For the \ac{RD} era, the conformal Hubble parameter $\mathcal{H}=a'/a=1/\eta$. The source term $\mathcal{S}^{(3)}_{l m}$ takes the form
\begin{equation}\label{eq:S}
	S_{lm}^{(3)}=S_{lm,1}^{(3)}+S_{lm,2}^{(3)}+S_{lm,3}^{(3)}+S_{lm,4}^{(3)} \ ,
\end{equation}
where $S_{lm,1}^{(3)}$, $S_{lm,2}^{(3)}$, $S_{lm,3}^{(3)}$ , and $S_{lm,4}^{(3)}$ are source terms of first order scalar perturbation, second order tensor perturbation, second order vector perturbation, and second order scalar perturbations, respectively. The explicit expressions of these source terms are given by
\begin{eqnarray}
		S_{lm,1}^{(3)}&=& \frac{1}{\mathcal{H}}\left(12 \mathcal{H} \phi^{(1)}-\phi^{(1)\prime}\right) \partial_l \phi^{(1)} \partial_m \phi^{(1)} \nonumber\\
		&-&\frac{1}{\mathcal{H}^3}\left(4 \mathcal{H} \phi^{(1)}-\phi^{(1)\prime}\right) \partial_l \phi^{(1)\prime} \partial_m \phi^{(1)\prime} \nonumber\\
		&+& \partial_l\left(\mathcal{H} \phi^{(1)}+\phi^{(1)\prime}\right) \partial_m\left(\mathcal{H} \phi^{(1)}+\phi^{(1)\prime}\right) \nonumber\\
		&\times&\frac{1}{3 \mathcal{H}^4}\left(2 \Delta \phi^{(1)}-9 \mathcal{H} \phi^{(1)\prime}\right) \ ,
	\label{eq:S1} 
\end{eqnarray}
\begin{eqnarray}
	S_{lm,2}^{(3)}&=&-\frac{1}{2}\phi^{(1)}\left( h_{lm}^{(2)''}+2 \mathcal{H}  h_{lm}^{(2)'}-\Delta h_{lm}^{(2)}\right) \nonumber\\
	&-&\phi^{(1)}\Delta h_{lm}^{(2)}-\phi^{(1)'}\mathcal{H}h_{lm}^{(2)}-\frac{1}{3}\Delta \phi^{(1)}h_{lm}^{(2)} \nonumber\\
	&-&\partial^b \phi^{(1)}\partial_b h_{lm}^{(2)} \ ,
	\label{eq:S2} 
\end{eqnarray}
\begin{eqnarray}
	S_{lm,3}^{(3)}&=&\phi^{(1)}\partial_l\left(V_m^{(2)'}+2 \mathcal{H}V_m^{(2)} \right)+2\phi^{(1)'}\partial_{(l}V_{m)}^{(2)}\nonumber\\
	&+&\phi^{(1)}\partial_m\left(V_l^{(2)'}+2 \mathcal{H}V_l^{(2)} \right)-\frac{\phi^{(1)}}{4\mathcal{H}}\partial_{(m}\Delta V_{l)}^{(2)}\nonumber\\
	&-&\frac{\phi^{(1)'}}{4\mathcal{H}^2}\partial_{(m}\Delta V_{l)}^{(2)}  \ ,
	\label{eq:S3} 
\end{eqnarray}
\begin{eqnarray}
	S_{lm,4}^{(3)}&=&\frac{1}{\mathcal{H}}\left(\phi^{(1)}\partial_l\partial_m\psi^{(2)'}+\phi^{(1)'}\partial_l\partial_m\phi^{(2)}\right.\nonumber\\
	&+&\left.\frac{1}{\mathcal{H}}\phi^{(1)'}\partial_l\partial_m\psi^{(2)'}\right)+3\phi^{(1)}\partial_l\partial_m\phi^{(2)} \ .
	\label{eq:S4} 
\end{eqnarray}
In Eq.~(\ref{eq:S3}), we used the symbol of symmetric tensor $T_{(lm)}\equiv\frac{1}{2}( T_{lm}+T_{ml})$. Eq.~(\ref{eq:h3}) in momentum space can be solved  by the Green’s function method, namely
\begin{eqnarray}
	h^{\lambda}_{\mathbf{k}}( \eta)=\frac{12}{k \eta} \int \mathrm{d} \tilde{\eta} \sin (k \eta-k \tilde{\eta}) \tilde{\eta}\mathcal{S}^{\lambda}_{\mathbf{k}}(\tilde{\eta}) \  ,
\end{eqnarray}
where we have defined $h^{\lambda}_{\mathbf{k}}( \eta)=\varepsilon^{\lambda, ij}(\mathbf{k})h_{ij}^{(3)}(\mathbf{k}, \eta)$ and  $\mathcal{S}^{\lambda}_{\mathbf{k}}( \eta)=-\varepsilon^{\lambda, lm}(\mathbf{k})S_{lm}^{(3)}(\mathbf{k}, \eta)$. And $\varepsilon_{i j}^{\lambda}(\mathbf{k})$ is the polarization tensor, which satisfies $\varepsilon_{i j}^{\lambda}(\mathbf{k}) \varepsilon^{\bar{\lambda}, ij}(\mathbf{k})\delta^{\lambda \bar{\lambda}}=2$ and $\delta^{\lambda\bar{\lambda}}\varepsilon^{\bar{\lambda}, lm}(\mathbf{k})\varepsilon^{\bar{\lambda}}{ij}(\mathbf{k})=\Lambda_{i j}^{lm}(\mathbf{k})+A_{i j}^{lm}(\mathbf{k})$. Here, $A_{i j}^{lm}(\mathbf{k})$ is defined as \cite{Chang:2020tji}
\begin{equation}\label{eq:A}
A_{i j}^{lm}\equiv\frac{1}{\sqrt{2}}\left(\bar{e}_ie_j-e_i\bar{e}_j \right)\frac{1}{\sqrt{2}}\left(\bar{e}_me_l-e_m\bar{e}_l \right) \  ,
\end{equation}
where $\left(\mathbf{k}_i/|k|,e_i,\bar{e}_i  \right)$ is the normalized bases in three dimensional momentum space. Note that $A_{i j}^{lm}(\mathbf{k})$ is an antisymmetric tensor, therefore, $A_{i j}^{lm}T_{lm}=A_{i j}^{lm}T^{ij}=0$ for arbitrary symmetric tensor $T_{ij}$. In the calculations of the second and the third order \acp{SIGW}, $A_{i j}^{lm}$ only acts on symmetric tensor. In this case, we obtain $A_{i j}^{lm}=0$ and $\delta^{\lambda\bar{\lambda}}\varepsilon^{\bar{\lambda}, lm}(\mathbf{k})\varepsilon^{\bar{\lambda}}{ij}(\mathbf{k})=\Lambda_{i j}^{lm}(\mathbf{k})$. There are four kinds of source terms in Eq.~(\ref{eq:h3}), it is convenient to divide $h^{\lambda}_{\mathbf{k}}( \eta)$ into four parts $h^{\lambda}_{\mathbf{k}}( \eta)=\sum_{i=1}^{4}h^{\lambda}_{\mathbf{k},i}( \eta)$. The formal expressions of $h^{\lambda}_{\mathbf{k},i}( \eta)$ are given by 
\begin{eqnarray}
	h^{\lambda}_{\mathbf{k},1}(\eta)&=&\int\frac{d^3p}{(2\pi)^{3/2}}\int\frac{d^3q}{(2\pi)^{3/2}}\varepsilon^{\lambda,lm}(\mathbf{k})(p_l-q_l)q_m \nonumber\\
	& &\times I_1^{(3)}(|\mathbf{k}-\mathbf{p}|,|\mathbf{p}-\mathbf{q}|,\mathbf{q},\eta)\zeta_{\mathbf{k}-\mathbf{p}} \zeta_{\mathbf{p}-\mathbf{q}} \zeta_{\mathbf{q}} \  ,
	\label{eq:H1}\\
	h^{\lambda}_{\mathbf{k},2}(\eta)&=&\int\frac{d^3p}{(2\pi)^{3/2}}\int\frac{d^3q}{(2\pi)^{3/2}}\varepsilon^{\lambda, lm}(\mathbf{k})\Lambda_{lm}^{ rs}(\mathbf{p})q_rq_s \nonumber\\
	& &\times
	I^{(3)}_2(|\mathbf{k}-\mathbf{p}|,|\mathbf{p}-\mathbf{q}|,\mathbf{q},\eta)\zeta_{\mathbf{k}-\mathbf{p}} \zeta_{\mathbf{p}-\mathbf{q}} \zeta_{\mathbf{q}} \ ,
	\label{eq:H2} \\
	h^{\lambda}_{\mathbf{k},3}(\eta)&=&\int\frac{d^3p}{(2\pi)^{3/2}}\int\frac{d^3q}{(2\pi)^{3/2}}\varepsilon^{\lambda,lm}(\mathbf{k})\mathcal{T}^r_{(m}(\textbf{p}) p_{l)}\frac{2p^s}{p^2}q_rq_s  \nonumber\\
	& &\times I^{(3)}_3(|\mathbf{k}-\mathbf{p}|,|\mathbf{p}-\mathbf{q}|,\mathbf{q},\eta)\zeta_{\mathbf{k}-\mathbf{p}}\zeta_{\mathbf{p}-\mathbf{q}} \zeta_{\mathbf{q}} \ ,
	\label{eq:H3}\\
	h^{\lambda}_{\mathbf{k},4}(\eta)&=&\int\frac{d^3p}{(2\pi)^{3/2}}\int\frac{d^3q}{(2\pi)^{3/2}}\varepsilon^{\lambda,lm}(\mathbf{k})p_lp_m \nonumber\\
	& &\times
	I^{(3)}_4(|\mathbf{k}-\mathbf{p}|,|\mathbf{p}-\mathbf{q}|,\mathbf{q},\eta)\zeta_{\mathbf{k}-\mathbf{p}} \zeta_{\mathbf{p}-\mathbf{q}} \zeta_{\mathbf{q}} \ ,
	\label{eq:H4}
\end{eqnarray}
where $I^{(3)}_i$ $(i=1,2,3,4)$ are kernel functions of third order \ac{SIGW}, which take the form of
\begin{eqnarray}
	I_{i}^{(3)}(u,\bar{u},\bar{v},x)&&=\frac{12}{k^{2}} \int_{0}^{x} \mathrm{d}\bar{x} \left(\frac{\bar{x}}{x} \sin(x-\bar{x}) f_{i}^{(3)}(u,\bar{u},\bar{v},\bar{x})\right) \ , \nonumber\\
	&&(i=1,2,3,4) \ .
	\label{eq:li}
\end{eqnarray}
Here, we have defined $|\mathbf{k}-\mathbf{p}|=uk$, $|\mathbf{k}-\mathbf{q}|=wk$,  $|\mathbf{p}-\mathbf{q}|=\bar{u}p=\bar{u}vk$, and $q=\bar{v}p=\bar{v}vk$. $f_{i}^{(3)}\left(u,\bar{u},\bar{v},\bar{x})\right)$$(i=1,2,3,4)$ are  transfer functions of four kinds of source terms in Eq.~(\ref{eq:S1})--Eq.~(\ref{eq:S4}), which are given in Appendix A. \ref{sec:A} We consider a monochromatic primordial power spectrum
\begin{eqnarray}\label{eq:mp}
	\mathcal{P}_{\zeta}=A_{\zeta}k_*\delta\left( k-k_* \right) \ ,
	\label{eq:P0}
\end{eqnarray}
where $k_*$  is the wavenumber at which the power spectrum has a $\delta$-function peak. In the case of a monochromatic primordial power spectrum, the explicit expression of the energy density spectra of third order \acp{SIGW} in the \ac{RD} era is given by \cite{Zhou:2021vcw}
\begin{eqnarray}
\Omega_{\mathrm{GW}}^{(3)}(\eta, k)=\frac{\rho^{(3)}_{\mathrm{GW}}(\eta, k)}{\rho_{\mathrm{tot}}(\eta)}=\frac{x^2}{216} \sum_{i,j=1}^{4}\mathcal{P}^{ij}_h \ ,
\end{eqnarray}
where $\mathcal{P}^{ij}_h$ is the power spectra of third order \acp{SIGW} 
\begin{eqnarray}
	\mathcal{P}^{i  j}_h&&=\frac{\mathcal{A}_{\zeta}^3 \tilde{k}^3}{2 \pi}	\Theta (3 - \tilde{k})\int_{\left| 1 - \frac{1}{\tilde{k}}		\right|}^{\min \left\{ \frac{2}{\tilde{k}}, 1 + \frac{1}{\tilde{k}}		\right\}}   {\rm d} v  \int_{w_-}^{w_+} {\rm d} w \nonumber\\
	& &\left(\frac{ v w}{\sqrt{Y (1 - X^2)}}\ I^{(3)}_i\left(u,v,\bar{u},\bar{v},x\right)  \right. ~ \ \ \ \ \ , \nonumber\\
	& &\left. \times  \ \sum^{3}_{a=1} \mathbb{P}^{i  j}_{a} \  I^{(3),a}_j\left( u',v',\bar{u}',\bar{v}',\eta \right) \ \right)_{u=\frac{1}{\tilde{k}}, \bar{u}=\bar{v}=\frac{1}{v \tilde{k}}}  
	\label{eq:Pij}
\end{eqnarray}
where $\tilde{k}=k/k_*$ is a dimensionless parameter. The explicit expressions of $X$, $Y$, and $w_{\pm}$ are given in Appendix B.\ref{sec:B} The details of polynomials $\mathbb{P}^{i  j}_{a}$ can be found in Ref.~\cite{Zhou:2021vcw}. Note that the energy density of \acp{GW} decays as radiation, then the current total energy density spectra of \acp{SIGW} can be approximated by \cite{Espinosa:2017sgp}
\begin{equation}\label{eq:tot}
	\begin{aligned}
		\Omega_{\mathrm{GW}}(\eta_0, k)\simeq \Omega_r \times \left(\Omega^{(2)}_{\mathrm{GW}}(\eta, k)+\Omega^{(3)}_{\mathrm{GW}}(\eta, k)\right) \ ,
	\end{aligned}
\end{equation}
where $\Omega^{(2)}_{\mathrm{GW}}(\eta, k)$ is the energy density spectrum of second order \acp{SIGW} which has been studied systematically in previous work \cite{Kohri:2018awv,Espinosa:2018eve}.

\begin{figure}
	\includegraphics[scale = 0.45]{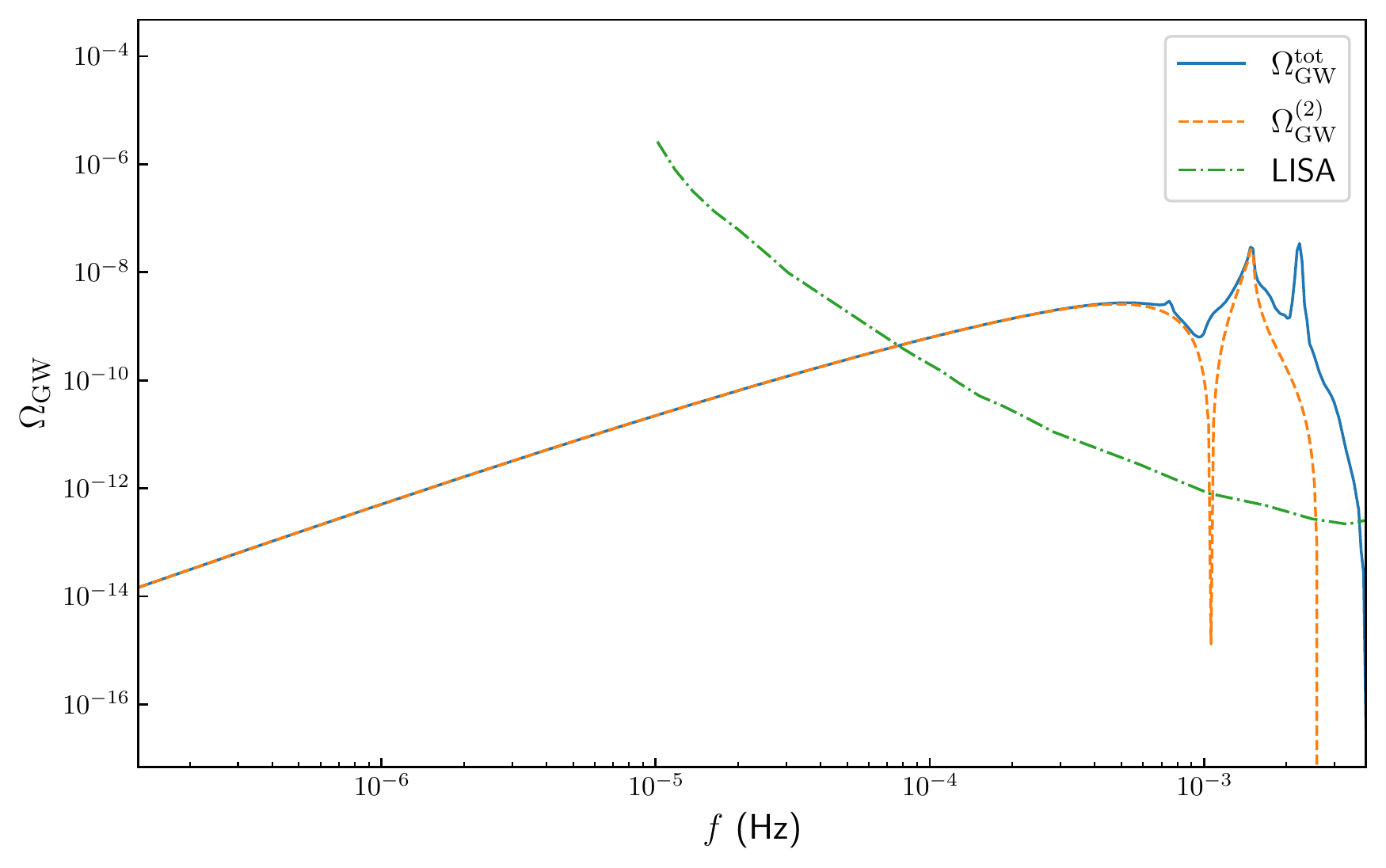}
	\caption{The energy density spectra of \acp{SIGW} to the second order $\Omega^{(2)}_{GW}$ (orange curve) and to the third order $\Omega^{\mathrm{tot}}_{GW}=\Omega^{(2)}_{GW}+\Omega^{(3)}_{GW}$ (blue curve) as function of frequency $f$. We have set $f_*=1.3\times 10^{-3}$Hz and $A_{\zeta}=0.02$.}\label{fig:ps}
\end{figure}

\emph{\acp{SIGW} as a probe of \acp{PBH}}---The monochromatic primordial power spectrum corresponds to a monochromatic \ac{PBH} formation. The possibility of forming a \acp{PBH} can be calculated in terms of the amplitude of primordial power spectrum \cite{Carr:2016drx,Nakama:2016gzw}
\begin{equation}
	\beta=\int_{\zeta_c}^{+\infty} \frac{\mathrm{d} \zeta}{\sqrt{2 \pi} \sigma} e^{-\zeta^2 / 2 \sigma^2}=\frac{1}{2} \operatorname{erfc}\left(\frac{\zeta_c}{\sqrt{2 A_{\zeta}}}\right) \ , \label{eq:1}
\end{equation}
 where $\sigma^2 \equiv\left\langle\zeta^2\right\rangle=\int \mathcal{P}_\zeta(k) \mathrm{d} \ln k=A_{\zeta}$ is the variance of the primordial curvature perturbation and $\zeta_c \simeq 1$ is the threshold value to form a \ac{PBH} \cite{Harada:2013epa,Musco:2008hv,Musco:2012au,Yuan:2019udt}. Moreover, the mass of the \ac{PBH} can be approximated by the frequency of the $\delta$-function peak
\begin{eqnarray}
	\frac{m_{\mathrm{pbh}}}{M_{\odot}} \approx 2.3 \times 10^{18}\left(\frac{3.91}{g_*^{\text {form }}}\right)^{1 / 6}\left(\frac{H_0}{f_*}\right)^2 \ , \label{eq:2}
\end{eqnarray}
where $f_*=k_*/(2\pi)$ and $g_*^{\text {form }}$ is the effective degrees of freedom when \acp{PBH} are formed. Then, the fraction of \acp{PBH} can be evaluated in terms of $\beta$ and $m_{\mathrm{pbh}}$
\begin{equation}
	f_{\mathrm{pbh}} \simeq 2.5 \times 10^8 \beta\left(\frac{g_*^{\text {form }}}{10.75}\right)^{-\frac{1}{4}}\left(\frac{m_{\mathrm{pbh}}}{M_{\odot}}\right)^{-\frac{1}{2}} \ . \label{eq:3}
\end{equation}
In the case of the monochromatic \ac{PBH} formation, Eq.~({\ref{eq:1}})--Eq.~(\ref{eq:3}) show that $\beta$ and $m_{\mathrm{pbh}}$ are determined by the amplitude of primordial power spectrum $A_{\zeta}$ and the frequency of the $\delta$-function peak $f_*$, respectively. Besides, the fraction of \acp{PBH} $f_{\mathrm{pbh}}$ increase with $k_*$ and $A_{\zeta}$. This strong correlation between \acp{PBH} and the concomitant \acp{SIGW} signals could be a promising approach to detecting \acp{PBH} in the upcoming \ac{GW} experiments, such as \ac{LISA}. Fig.~\ref{fig:ps} shows the energy density spectra of \acp{SIGW} compared with the sensitivity curves of \ac{LISA}. We find that the effects of the third order \acp{SIGW} extend the cutoff frequency from $2f_*$ to $3f_*$ and have significant observational implications for \ac{LISA}.

\begin{figure}
	\includegraphics[scale = 0.45]{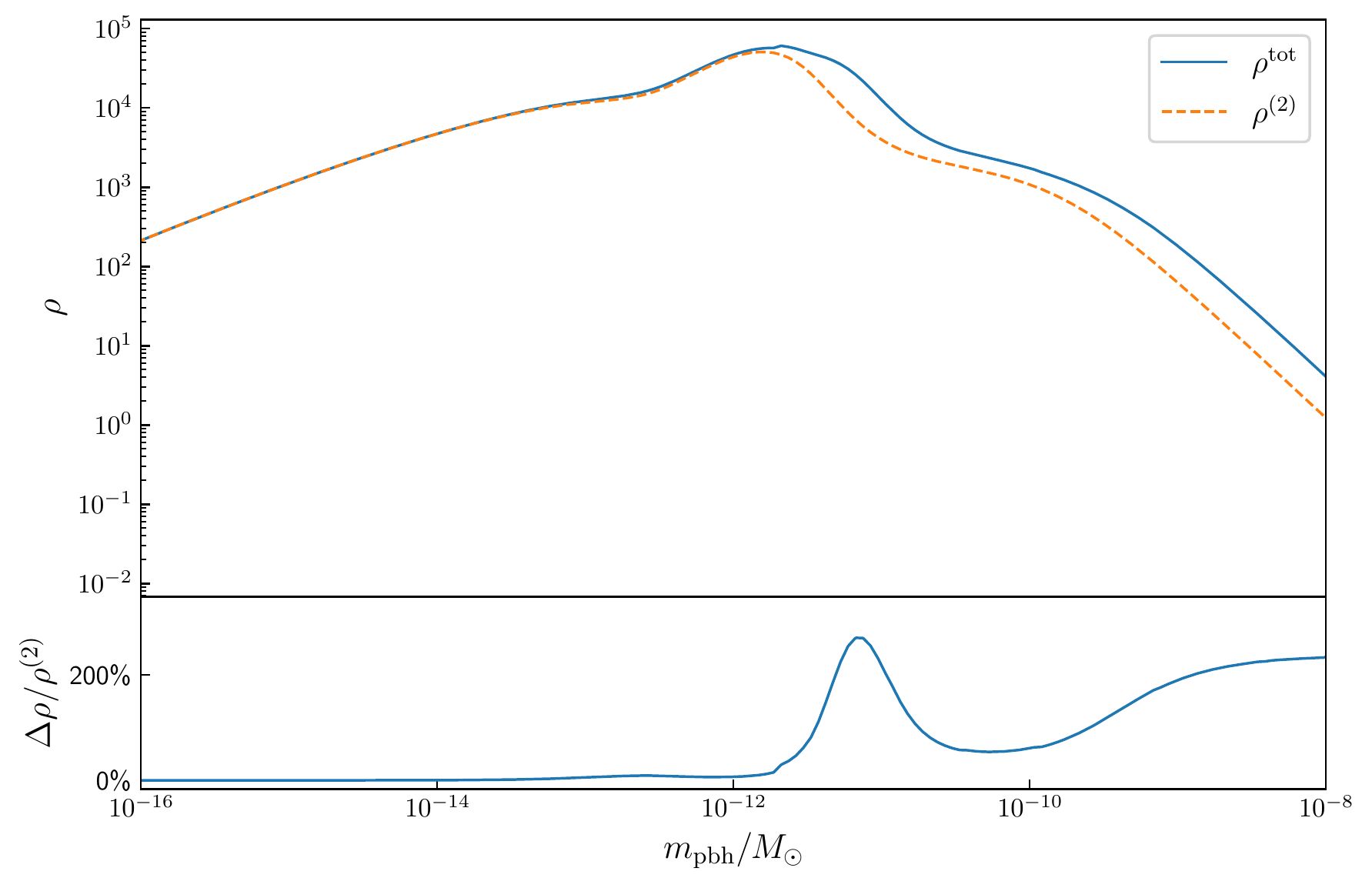}
	\caption{The SNR of LISA as a function of $m_{\mathrm{pbh}}$ for $\Omega^{\mathrm{tot}}_{GW}=\Omega^{(2)}_{GW}+\Omega^{(3)}_{GW}$ ($\rho^\mathrm{tot}=\rho^{(2)}+\rho^{(3)}$, blue solid curve) and $\Omega^{(2)}_{GW}$ ($\rho^{(2)}$, orange dashed curve). We have set $A_{\zeta}=0.02$. We also give $\Delta \rho/\rho^{tot} = \rho^\mathrm{tot}/\rho^{(2)}-1$ in the bottom panel. }\label{fig:SNR1}
\end{figure}

\begin{figure}
	\includegraphics[scale = 0.45]{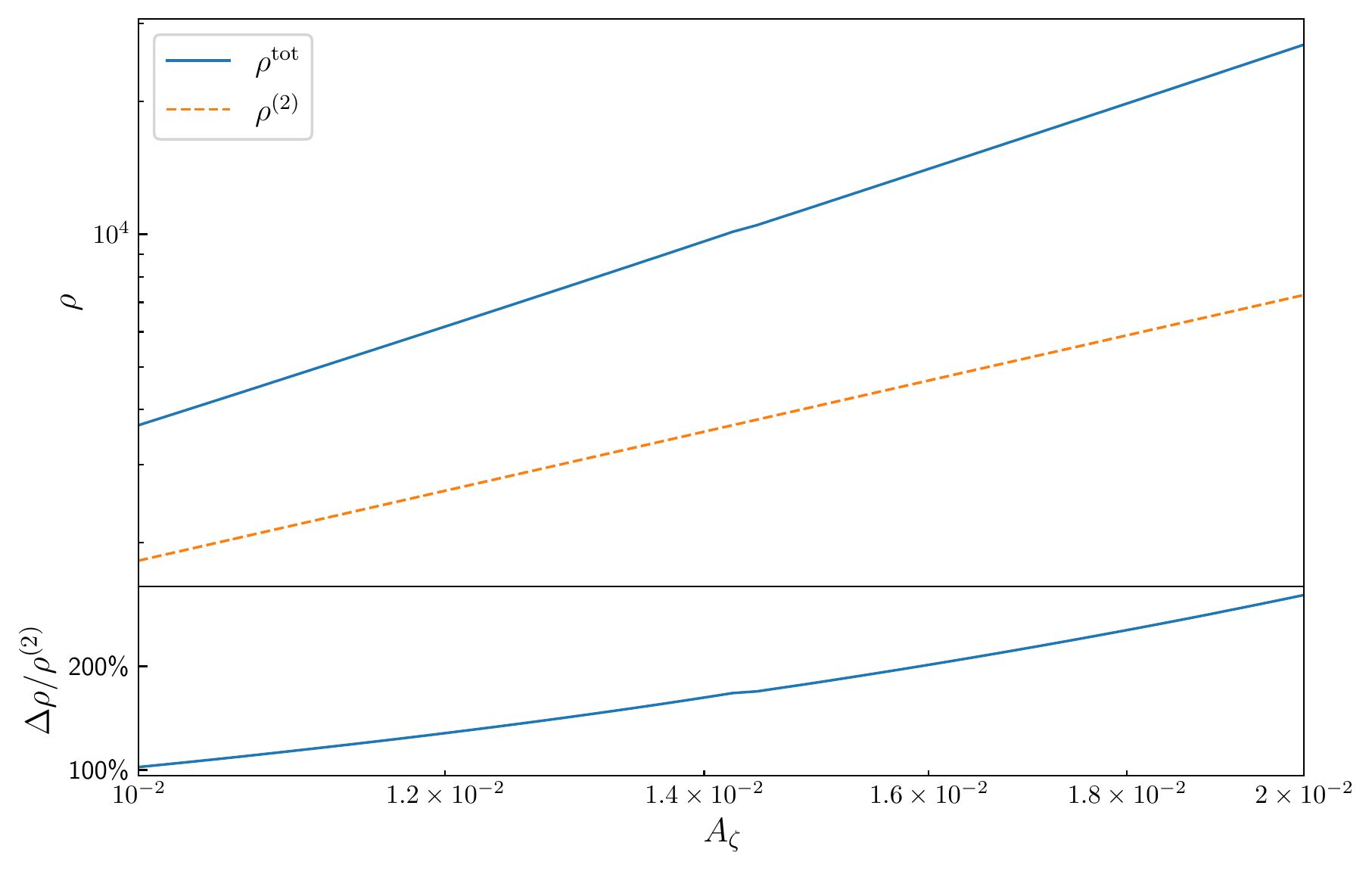}
	\caption{The SNR of LISA as a function of $A_{\zeta}$ for $\Omega^{\mathrm{tot}}_{GW}=\Omega^{(2)}_{GW}+\Omega^{(3)}_{GW}$ ($\rho^\mathrm{tot}=\rho^{(2)}+\rho^{(3)}$, blue solid curve) and $\Omega^{(2)}_{GW}$ ($\rho^{(2)}$, orange dashed curve). We have set $f_*=1.3\times10^{-3}$. We also give $\Delta \rho/\rho^{tot} = \rho^\mathrm{tot}/\rho^{(2)}-1$ in the bottom panel. }\label{fig:SNR2}
\end{figure}

To quantify the effects of the third order \acp{SIGW}, we calculate the SNR $\rho$ for \ac{LISA}, which is given by \cite{Thrane:2013oya,Thrane:2013oya,Robson:2018ifk,Kuroda:2015owv,Kuroda:2015owv,Siemens:2013zla,Zhao:2022kvz}
\begin{equation}
	\rho=\sqrt{T}\left[\int \mathrm{d} f\left(\frac{\Omega_{\mathrm{GW}}(f)}{\Omega_n(f)}\right)^2\right]^{1 / 2} \ ,
\end{equation}
where $\Omega_n(f) = 2\i^2f^3S_n/3H_0^2$ and $S_n$ is the strain noise power spectral density, $T$ is the observation time. Here, we set $T=4$ years. In Fig.~\ref{fig:SNR1} and Fig.~\ref{fig:SNR2}, we plot the \ac{SNR} curves obtained for LISA experiment. More precisely, Fig.~\ref{fig:SNR1} shows the relation between \ac{SNR} and $m_{\mathrm{pbh}}$ for a given $A_{\zeta}$. It shows that the effects of third order \acp{SIGW} lead to about $200\%$ increase of the \ac{SNR} for frequency band from $1.6\times 10^{-3}$Hz to $10^{-5}$Hz corresponding to \acp{PBH} with mass range $4\times 10^{-12}M_{\odot} \sim 10^{-7}M_{\odot}$. Fig.~\ref{fig:SNR2} shows the the relation between \ac{SNR} and $A_{\zeta}$ for a given $m_{\mathrm{pbh}}$. Obviously, \ac{SNR} and corresponding third order correction increase with the amplitude of the small-scale primordial power spectra $A_{\zeta}$.
Since the effects of the third order \acp{SIGW} significantly affect the total energy density spectrum of \acp{SIGW}, it is necessary to compare the energy density of the third order \acp{SIGW} $\rho_{GW}^{(3)}$ with the energy density of the second order \acp{SIGW} $\rho_{GW}^{(2)}$.  If we assume $\rho_{GW}^{(3)}<\rho_{GW}^{(2)}$, we will obtain a critical value $A_*$ of the amplitude of the small-scale primordial power spectra. The critical value $A_*$ can be calculated by
\begin{equation}\label{eq:23}
	\int_{0}^{2k_*} \Omega^{(2)}_{\mathrm{GW}}(\eta, k)\mathrm{d}\ln k=\int_{0}^{3k_*} \Omega^{(3)}_{\mathrm{GW}}(\eta, k)\mathrm{d}\ln k \ .
\end{equation}
Solving Eq.~(\ref{eq:23}), we obtain $A_*=1.76\times 10^{-2}$. If $A_{\zeta}>A_*$, then the energy density of third order \acp{SIGW} will be larger than the energy density of second order \acp{SIGW}. Note that this critical value $A_*$ is just an upper limit, because $\Omega^{(3)}_{\mathrm{GW}}(\eta, k)$ not only comes from $\langle h^{(3)}_{\mathbf{k}} h^{(3)}_{\mathbf{k}'}\rangle$, but also comes from $\langle h^{(4)}_{\mathbf{k}} h^{(2)}_{\mathbf{k}'}\rangle$.  The study of fourth order \acp{SIGW} is extremely difficult since one has to calculate all three kinds of third order perturbations firstly. In this paper, we do not consider the effect of $\langle h^{(4)}_{\mathbf{k}} h^{(2)}_{\mathbf{k}'}\rangle$. Predictably, it will lead to a larger \ac{SNR} and a smaller critical value $A_*$.

\emph{Conclusion and discussion}---In this paper, we studied the third order \acp{SIGW} and calculated corresponding energy density spectrum. We found that the third order correction extend the cutoff frequency from $2f_*$ to $3f_*$. The relations between \ac{SNR} of
\ac{LISA} and the monochromatic primordial power spectrum were studied systematically. For \acp{PBH} with mass range $4\times 10^{-12}M_{\odot} \sim 10^{-7}M_{\odot}$, the effects of third order \acp{SIGW} lead to around $200\%$ increase of the \ac{SNR}. For a given $f_*$, \ac{SNR} and third order correction increase with $A_{\zeta}$. We pointed that there exists a critical value $A_*$, such that when $A_{\zeta}>A_*$, the energy density of third order \acp{SIGW} will be larger than the energy density of second order \acp{SIGW}. We conclude that the third order SIGWs are dominated by the source term of second order scalar perturbation $\mathcal{S}^{(3)}_{4}\sim \phi^{(1)}\psi^{(2)}$. The third order SIGWs from a general primordial power spectrum were not considered here because of the difficulties of numerical calculation. The complete study of higher order SIGWs might be presented in the future work.

\vspace{0.3cm}
\begin{acknowledgements} 
We thank Prof.~S. Wang, Dr.~Y.H. Yu, and Dr.~ J.P. Li for the useful discussions. This work has been funded by the National Nature Science Foundation of China under grant No. 12075249 and 11690022, and the Key Research Program of the Chinese Academy of Sciences under Grant No. XDPB15. We acknowledge the \texttt{xPand} package \cite{Pitrou:2013hga}. 
\end{acknowledgements}

\bibliography{PBHthird}

\begin{thebibliography}{106}%
\makeatletter
\providecommand \@ifxundefined [1]{%
 \@ifx{#1\undefined}
}%
\providecommand \@ifnum [1]{%
 \ifnum #1\expandafter \@firstoftwo
 \else \expandafter \@secondoftwo
 \fi
}%
\providecommand \@ifx [1]{%
 \ifx #1\expandafter \@firstoftwo
 \else \expandafter \@secondoftwo
 \fi
}%
\providecommand \natexlab [1]{#1}%
\providecommand \enquote  [1]{``#1''}%
\providecommand \bibnamefont  [1]{#1}%
\providecommand \bibfnamefont [1]{#1}%
\providecommand \citenamefont [1]{#1}%
\providecommand \href@noop [0]{\@secondoftwo}%
\providecommand \href [0]{\begingroup \@sanitize@url \@href}%
\providecommand \@href[1]{\@@startlink{#1}\@@href}%
\providecommand \@@href[1]{\endgroup#1\@@endlink}%
\providecommand \@sanitize@url [0]{\catcode `\\12\catcode `\$12\catcode
  `\&12\catcode `\#12\catcode `\^12\catcode `\_12\catcode `\%12\relax}%
\providecommand \@@startlink[1]{}%
\providecommand \@@endlink[0]{}%
\providecommand \url  [0]{\begingroup\@sanitize@url \@url }%
\providecommand \@url [1]{\endgroup\@href {#1}{\urlprefix }}%
\providecommand \urlprefix  [0]{URL }%
\providecommand \Eprint [0]{\href }%
\providecommand \doibase [0]{http://dx.doi.org/}%
\providecommand \selectlanguage [0]{\@gobble}%
\providecommand \bibinfo  [0]{\@secondoftwo}%
\providecommand \bibfield  [0]{\@secondoftwo}%
\providecommand \translation [1]{[#1]}%
\providecommand \BibitemOpen [0]{}%
\providecommand \bibitemStop [0]{}%
\providecommand \bibitemNoStop [0]{.\EOS\space}%
\providecommand \EOS [0]{\spacefactor3000\relax}%
\providecommand \BibitemShut  [1]{\csname bibitem#1\endcsname}%
\let\auto@bib@innerbib\@empty
\bibitem [{\citenamefont {Aghanim}\ \emph {et~al.}(2020)\citenamefont {Aghanim}
  \emph {et~al.}}]{Planck:2018vyg}%
  \BibitemOpen
  \bibfield  {author} {\bibinfo {author} {\bibfnamefont {N.}~\bibnamefont
  {Aghanim}} \emph {et~al.} (\bibinfo {collaboration} {Planck}),\ }\href
  {\doibase 10.1051/0004-6361/201833910} {\bibfield  {journal} {\bibinfo
  {journal} {Astron. Astrophys.}\ }\textbf {\bibinfo {volume} {641}},\ \bibinfo
  {pages} {A6} (\bibinfo {year} {2020})},\ \bibinfo {note} {[Erratum:
  Astron.Astrophys. 652, C4 (2021)]},\ \Eprint
  {http://arxiv.org/abs/1807.06209} {arXiv:1807.06209 [astro-ph.CO]}
  \BibitemShut {NoStop}%
\bibitem [{\citenamefont {Abdalla}\ \emph {et~al.}(2022)\citenamefont {Abdalla}
  \emph {et~al.}}]{Abdalla:2022yfr}%
  \BibitemOpen
  \bibfield  {author} {\bibinfo {author} {\bibfnamefont {E.}~\bibnamefont
  {Abdalla}} \emph {et~al.},\ }\href {\doibase 10.1016/j.jheap.2022.04.002}
  {\bibfield  {journal} {\bibinfo  {journal} {JHEAp}\ }\textbf {\bibinfo
  {volume} {34}},\ \bibinfo {pages} {49} (\bibinfo {year} {2022})},\ \Eprint
  {http://arxiv.org/abs/2203.06142} {arXiv:2203.06142 [astro-ph.CO]}
  \BibitemShut {NoStop}%
\bibitem [{\citenamefont {Bringmann}\ \emph {et~al.}(2012)\citenamefont
  {Bringmann}, \citenamefont {Scott},\ and\ \citenamefont
  {Akrami}}]{Bringmann:2011ut}%
  \BibitemOpen
  \bibfield  {author} {\bibinfo {author} {\bibfnamefont {T.}~\bibnamefont
  {Bringmann}}, \bibinfo {author} {\bibfnamefont {P.}~\bibnamefont {Scott}}, \
  and\ \bibinfo {author} {\bibfnamefont {Y.}~\bibnamefont {Akrami}},\ }\href
  {\doibase 10.1103/PhysRevD.85.125027} {\bibfield  {journal} {\bibinfo
  {journal} {Phys. Rev. D}\ }\textbf {\bibinfo {volume} {85}},\ \bibinfo
  {pages} {125027} (\bibinfo {year} {2012})},\ \Eprint
  {http://arxiv.org/abs/1110.2484} {arXiv:1110.2484 [astro-ph.CO]} \BibitemShut
  {NoStop}%
\bibitem [{\citenamefont {Sasaki}\ \emph {et~al.}(2018)\citenamefont {Sasaki},
  \citenamefont {Suyama}, \citenamefont {Tanaka},\ and\ \citenamefont
  {Yokoyama}}]{Sasaki:2018dmp}%
  \BibitemOpen
  \bibfield  {author} {\bibinfo {author} {\bibfnamefont {M.}~\bibnamefont
  {Sasaki}}, \bibinfo {author} {\bibfnamefont {T.}~\bibnamefont {Suyama}},
  \bibinfo {author} {\bibfnamefont {T.}~\bibnamefont {Tanaka}}, \ and\ \bibinfo
  {author} {\bibfnamefont {S.}~\bibnamefont {Yokoyama}},\ }\href {\doibase
  10.1088/1361-6382/aaa7b4} {\bibfield  {journal} {\bibinfo  {journal} {Class.
  Quant. Grav.}\ }\textbf {\bibinfo {volume} {35}},\ \bibinfo {pages} {063001}
  (\bibinfo {year} {2018})},\ \Eprint {http://arxiv.org/abs/1801.05235}
  {arXiv:1801.05235 [astro-ph.CO]} \BibitemShut {NoStop}%
\bibitem [{\citenamefont {Carr}(1975)}]{Carr:1975qj}%
  \BibitemOpen
  \bibfield  {author} {\bibinfo {author} {\bibfnamefont {B.~J.}\ \bibnamefont
  {Carr}},\ }\href {\doibase 10.1086/153853} {\bibfield  {journal} {\bibinfo
  {journal} {Astrophys. J.}\ }\textbf {\bibinfo {volume} {201}},\ \bibinfo
  {pages} {1} (\bibinfo {year} {1975})}\BibitemShut {NoStop}%
\bibitem [{\citenamefont {Carr}\ and\ \citenamefont
  {Lidsey}(1993)}]{Carr:1993aq}%
  \BibitemOpen
  \bibfield  {author} {\bibinfo {author} {\bibfnamefont {B.~J.}\ \bibnamefont
  {Carr}}\ and\ \bibinfo {author} {\bibfnamefont {J.~E.}\ \bibnamefont
  {Lidsey}},\ }\href {\doibase 10.1103/PhysRevD.48.543} {\bibfield  {journal}
  {\bibinfo  {journal} {Phys. Rev. D}\ }\textbf {\bibinfo {volume} {48}},\
  \bibinfo {pages} {543} (\bibinfo {year} {1993})}\BibitemShut {NoStop}%
\bibitem [{\citenamefont {Carr}\ \emph {et~al.}(2016)\citenamefont {Carr},
  \citenamefont {Kuhnel},\ and\ \citenamefont {Sandstad}}]{Carr:2016drx}%
  \BibitemOpen
  \bibfield  {author} {\bibinfo {author} {\bibfnamefont {B.}~\bibnamefont
  {Carr}}, \bibinfo {author} {\bibfnamefont {F.}~\bibnamefont {Kuhnel}}, \ and\
  \bibinfo {author} {\bibfnamefont {M.}~\bibnamefont {Sandstad}},\ }\href
  {\doibase 10.1103/PhysRevD.94.083504} {\bibfield  {journal} {\bibinfo
  {journal} {Phys. Rev. D}\ }\textbf {\bibinfo {volume} {94}},\ \bibinfo
  {pages} {083504} (\bibinfo {year} {2016})},\ \Eprint
  {http://arxiv.org/abs/1607.06077} {arXiv:1607.06077 [astro-ph.CO]}
  \BibitemShut {NoStop}%
\bibitem [{\citenamefont {Carr}\ \emph {et~al.}(2021)\citenamefont {Carr},
  \citenamefont {Kohri}, \citenamefont {Sendouda},\ and\ \citenamefont
  {Yokoyama}}]{Carr:2020gox}%
  \BibitemOpen
  \bibfield  {author} {\bibinfo {author} {\bibfnamefont {B.}~\bibnamefont
  {Carr}}, \bibinfo {author} {\bibfnamefont {K.}~\bibnamefont {Kohri}},
  \bibinfo {author} {\bibfnamefont {Y.}~\bibnamefont {Sendouda}}, \ and\
  \bibinfo {author} {\bibfnamefont {J.}~\bibnamefont {Yokoyama}},\ }\href
  {\doibase 10.1088/1361-6633/ac1e31} {\bibfield  {journal} {\bibinfo
  {journal} {Rept. Prog. Phys.}\ }\textbf {\bibinfo {volume} {84}},\ \bibinfo
  {pages} {116902} (\bibinfo {year} {2021})},\ \Eprint
  {http://arxiv.org/abs/2002.12778} {arXiv:2002.12778 [astro-ph.CO]}
  \BibitemShut {NoStop}%
\bibitem [{\citenamefont {Carr}\ and\ \citenamefont
  {Kuhnel}(2020)}]{Carr:2020xqk}%
  \BibitemOpen
  \bibfield  {author} {\bibinfo {author} {\bibfnamefont {B.}~\bibnamefont
  {Carr}}\ and\ \bibinfo {author} {\bibfnamefont {F.}~\bibnamefont {Kuhnel}},\
  }\href {\doibase 10.1146/annurev-nucl-050520-125911} {\bibfield  {journal}
  {\bibinfo  {journal} {Ann. Rev. Nucl. Part. Sci.}\ }\textbf {\bibinfo
  {volume} {70}},\ \bibinfo {pages} {355} (\bibinfo {year} {2020})},\ \Eprint
  {http://arxiv.org/abs/2006.02838} {arXiv:2006.02838 [astro-ph.CO]}
  \BibitemShut {NoStop}%
\bibitem [{\citenamefont {Garcia-Bellido}\ and\ \citenamefont
  {Ruiz~Morales}(2017)}]{Garcia-Bellido:2017mdw}%
  \BibitemOpen
  \bibfield  {author} {\bibinfo {author} {\bibfnamefont {J.}~\bibnamefont
  {Garcia-Bellido}}\ and\ \bibinfo {author} {\bibfnamefont {E.}~\bibnamefont
  {Ruiz~Morales}},\ }\href {\doibase 10.1016/j.dark.2017.09.007} {\bibfield
  {journal} {\bibinfo  {journal} {Phys. Dark Univ.}\ }\textbf {\bibinfo
  {volume} {18}},\ \bibinfo {pages} {47} (\bibinfo {year} {2017})},\ \Eprint
  {http://arxiv.org/abs/1702.03901} {arXiv:1702.03901 [astro-ph.CO]}
  \BibitemShut {NoStop}%
\bibitem [{\citenamefont {Ivanov}\ \emph {et~al.}(1994)\citenamefont {Ivanov},
  \citenamefont {Naselsky},\ and\ \citenamefont {Novikov}}]{Ivanov:1994pa}%
  \BibitemOpen
  \bibfield  {author} {\bibinfo {author} {\bibfnamefont {P.}~\bibnamefont
  {Ivanov}}, \bibinfo {author} {\bibfnamefont {P.}~\bibnamefont {Naselsky}}, \
  and\ \bibinfo {author} {\bibfnamefont {I.}~\bibnamefont {Novikov}},\ }\href
  {\doibase 10.1103/PhysRevD.50.7173} {\bibfield  {journal} {\bibinfo
  {journal} {Phys. Rev. D}\ }\textbf {\bibinfo {volume} {50}},\ \bibinfo
  {pages} {7173} (\bibinfo {year} {1994})}\BibitemShut {NoStop}%
\bibitem [{\citenamefont {De~Luca}\ \emph {et~al.}(2021)\citenamefont
  {De~Luca}, \citenamefont {Franciolini},\ and\ \citenamefont
  {Riotto}}]{DeLuca:2020agl}%
  \BibitemOpen
  \bibfield  {author} {\bibinfo {author} {\bibfnamefont {V.}~\bibnamefont
  {De~Luca}}, \bibinfo {author} {\bibfnamefont {G.}~\bibnamefont
  {Franciolini}}, \ and\ \bibinfo {author} {\bibfnamefont {A.}~\bibnamefont
  {Riotto}},\ }\href {\doibase 10.1103/PhysRevLett.126.041303} {\bibfield
  {journal} {\bibinfo  {journal} {Phys. Rev. Lett.}\ }\textbf {\bibinfo
  {volume} {126}},\ \bibinfo {pages} {041303} (\bibinfo {year} {2021})},\
  \Eprint {http://arxiv.org/abs/2009.08268} {arXiv:2009.08268 [astro-ph.CO]}
  \BibitemShut {NoStop}%
\bibitem [{\citenamefont {De~Luca}\ \emph
  {et~al.}(2020{\natexlab{a}})\citenamefont {De~Luca}, \citenamefont
  {Franciolini}, \citenamefont {Pani},\ and\ \citenamefont
  {Riotto}}]{DeLuca:2020qqa}%
  \BibitemOpen
  \bibfield  {author} {\bibinfo {author} {\bibfnamefont {V.}~\bibnamefont
  {De~Luca}}, \bibinfo {author} {\bibfnamefont {G.}~\bibnamefont
  {Franciolini}}, \bibinfo {author} {\bibfnamefont {P.}~\bibnamefont {Pani}}, \
  and\ \bibinfo {author} {\bibfnamefont {A.}~\bibnamefont {Riotto}},\ }\href
  {\doibase 10.1088/1475-7516/2020/06/044} {\bibfield  {journal} {\bibinfo
  {journal} {JCAP}\ }\textbf {\bibinfo {volume} {06}},\ \bibinfo {pages} {044}
  (\bibinfo {year} {2020}{\natexlab{a}})},\ \Eprint
  {http://arxiv.org/abs/2005.05641} {arXiv:2005.05641 [astro-ph.CO]}
  \BibitemShut {NoStop}%
\bibitem [{\citenamefont {Vaskonen}\ and\ \citenamefont
  {Veerm\"ae}(2021)}]{Vaskonen:2020lbd}%
  \BibitemOpen
  \bibfield  {author} {\bibinfo {author} {\bibfnamefont {V.}~\bibnamefont
  {Vaskonen}}\ and\ \bibinfo {author} {\bibfnamefont {H.}~\bibnamefont
  {Veerm\"ae}},\ }\href {\doibase 10.1103/PhysRevLett.126.051303} {\bibfield
  {journal} {\bibinfo  {journal} {Phys. Rev. Lett.}\ }\textbf {\bibinfo
  {volume} {126}},\ \bibinfo {pages} {051303} (\bibinfo {year} {2021})},\
  \Eprint {http://arxiv.org/abs/2009.07832} {arXiv:2009.07832 [astro-ph.CO]}
  \BibitemShut {NoStop}%
\bibitem [{\citenamefont {Braglia}\ \emph {et~al.}(2020)\citenamefont
  {Braglia}, \citenamefont {Hazra}, \citenamefont {Finelli}, \citenamefont
  {Smoot}, \citenamefont {Sriramkumar},\ and\ \citenamefont
  {Starobinsky}}]{Braglia:2020eai}%
  \BibitemOpen
  \bibfield  {author} {\bibinfo {author} {\bibfnamefont {M.}~\bibnamefont
  {Braglia}}, \bibinfo {author} {\bibfnamefont {D.~K.}\ \bibnamefont {Hazra}},
  \bibinfo {author} {\bibfnamefont {F.}~\bibnamefont {Finelli}}, \bibinfo
  {author} {\bibfnamefont {G.~F.}\ \bibnamefont {Smoot}}, \bibinfo {author}
  {\bibfnamefont {L.}~\bibnamefont {Sriramkumar}}, \ and\ \bibinfo {author}
  {\bibfnamefont {A.~A.}\ \bibnamefont {Starobinsky}},\ }\href {\doibase
  10.1088/1475-7516/2020/08/001} {\bibfield  {journal} {\bibinfo  {journal}
  {JCAP}\ }\textbf {\bibinfo {volume} {08}},\ \bibinfo {pages} {001} (\bibinfo
  {year} {2020})},\ \Eprint {http://arxiv.org/abs/2005.02895} {arXiv:2005.02895
  [astro-ph.CO]} \BibitemShut {NoStop}%
\bibitem [{\citenamefont {Bartolo}\ \emph {et~al.}(2019)\citenamefont
  {Bartolo}, \citenamefont {De~Luca}, \citenamefont {Franciolini},
  \citenamefont {Lewis}, \citenamefont {Peloso},\ and\ \citenamefont
  {Riotto}}]{Bartolo:2018evs}%
  \BibitemOpen
  \bibfield  {author} {\bibinfo {author} {\bibfnamefont {N.}~\bibnamefont
  {Bartolo}}, \bibinfo {author} {\bibfnamefont {V.}~\bibnamefont {De~Luca}},
  \bibinfo {author} {\bibfnamefont {G.}~\bibnamefont {Franciolini}}, \bibinfo
  {author} {\bibfnamefont {A.}~\bibnamefont {Lewis}}, \bibinfo {author}
  {\bibfnamefont {M.}~\bibnamefont {Peloso}}, \ and\ \bibinfo {author}
  {\bibfnamefont {A.}~\bibnamefont {Riotto}},\ }\href {\doibase
  10.1103/PhysRevLett.122.211301} {\bibfield  {journal} {\bibinfo  {journal}
  {Phys. Rev. Lett.}\ }\textbf {\bibinfo {volume} {122}},\ \bibinfo {pages}
  {211301} (\bibinfo {year} {2019})},\ \Eprint
  {http://arxiv.org/abs/1810.12218} {arXiv:1810.12218 [astro-ph.CO]}
  \BibitemShut {NoStop}%
\bibitem [{\citenamefont {Byrnes}\ \emph {et~al.}(2018)\citenamefont {Byrnes},
  \citenamefont {Hindmarsh}, \citenamefont {Young},\ and\ \citenamefont
  {Hawkins}}]{Byrnes:2018clq}%
  \BibitemOpen
  \bibfield  {author} {\bibinfo {author} {\bibfnamefont {C.~T.}\ \bibnamefont
  {Byrnes}}, \bibinfo {author} {\bibfnamefont {M.}~\bibnamefont {Hindmarsh}},
  \bibinfo {author} {\bibfnamefont {S.}~\bibnamefont {Young}}, \ and\ \bibinfo
  {author} {\bibfnamefont {M.~R.~S.}\ \bibnamefont {Hawkins}},\ }\href
  {\doibase 10.1088/1475-7516/2018/08/041} {\bibfield  {journal} {\bibinfo
  {journal} {JCAP}\ }\textbf {\bibinfo {volume} {08}},\ \bibinfo {pages} {041}
  (\bibinfo {year} {2018})},\ \Eprint {http://arxiv.org/abs/1801.06138}
  {arXiv:1801.06138 [astro-ph.CO]} \BibitemShut {NoStop}%
\bibitem [{\citenamefont {Ballesteros}\ and\ \citenamefont
  {Taoso}(2018)}]{Ballesteros:2017fsr}%
  \BibitemOpen
  \bibfield  {author} {\bibinfo {author} {\bibfnamefont {G.}~\bibnamefont
  {Ballesteros}}\ and\ \bibinfo {author} {\bibfnamefont {M.}~\bibnamefont
  {Taoso}},\ }\href {\doibase 10.1103/PhysRevD.97.023501} {\bibfield  {journal}
  {\bibinfo  {journal} {Phys. Rev. D}\ }\textbf {\bibinfo {volume} {97}},\
  \bibinfo {pages} {023501} (\bibinfo {year} {2018})},\ \Eprint
  {http://arxiv.org/abs/1709.05565} {arXiv:1709.05565 [hep-ph]} \BibitemShut
  {NoStop}%
\bibitem [{\citenamefont {Ezquiaga}\ \emph {et~al.}(2018)\citenamefont
  {Ezquiaga}, \citenamefont {Garcia-Bellido},\ and\ \citenamefont
  {Ruiz~Morales}}]{Ezquiaga:2017fvi}%
  \BibitemOpen
  \bibfield  {author} {\bibinfo {author} {\bibfnamefont {J.~M.}\ \bibnamefont
  {Ezquiaga}}, \bibinfo {author} {\bibfnamefont {J.}~\bibnamefont
  {Garcia-Bellido}}, \ and\ \bibinfo {author} {\bibfnamefont {E.}~\bibnamefont
  {Ruiz~Morales}},\ }\href {\doibase 10.1016/j.physletb.2017.11.039} {\bibfield
   {journal} {\bibinfo  {journal} {Phys. Lett. B}\ }\textbf {\bibinfo {volume}
  {776}},\ \bibinfo {pages} {345} (\bibinfo {year} {2018})},\ \Eprint
  {http://arxiv.org/abs/1705.04861} {arXiv:1705.04861 [astro-ph.CO]}
  \BibitemShut {NoStop}%
\bibitem [{\citenamefont {Inomata}\ \emph {et~al.}(2017)\citenamefont
  {Inomata}, \citenamefont {Kawasaki}, \citenamefont {Mukaida}, \citenamefont
  {Tada},\ and\ \citenamefont {Yanagida}}]{Inomata:2016rbd}%
  \BibitemOpen
  \bibfield  {author} {\bibinfo {author} {\bibfnamefont {K.}~\bibnamefont
  {Inomata}}, \bibinfo {author} {\bibfnamefont {M.}~\bibnamefont {Kawasaki}},
  \bibinfo {author} {\bibfnamefont {K.}~\bibnamefont {Mukaida}}, \bibinfo
  {author} {\bibfnamefont {Y.}~\bibnamefont {Tada}}, \ and\ \bibinfo {author}
  {\bibfnamefont {T.~T.}\ \bibnamefont {Yanagida}},\ }\href {\doibase
  10.1103/PhysRevD.95.123510} {\bibfield  {journal} {\bibinfo  {journal} {Phys.
  Rev. D}\ }\textbf {\bibinfo {volume} {95}},\ \bibinfo {pages} {123510}
  (\bibinfo {year} {2017})},\ \Eprint {http://arxiv.org/abs/1611.06130}
  {arXiv:1611.06130 [astro-ph.CO]} \BibitemShut {NoStop}%
\bibitem [{\citenamefont {Kawasaki}\ \emph {et~al.}(2016)\citenamefont
  {Kawasaki}, \citenamefont {Kusenko}, \citenamefont {Tada},\ and\
  \citenamefont {Yanagida}}]{Kawasaki:2016pql}%
  \BibitemOpen
  \bibfield  {author} {\bibinfo {author} {\bibfnamefont {M.}~\bibnamefont
  {Kawasaki}}, \bibinfo {author} {\bibfnamefont {A.}~\bibnamefont {Kusenko}},
  \bibinfo {author} {\bibfnamefont {Y.}~\bibnamefont {Tada}}, \ and\ \bibinfo
  {author} {\bibfnamefont {T.~T.}\ \bibnamefont {Yanagida}},\ }\href {\doibase
  10.1103/PhysRevD.94.083523} {\bibfield  {journal} {\bibinfo  {journal} {Phys.
  Rev. D}\ }\textbf {\bibinfo {volume} {94}},\ \bibinfo {pages} {083523}
  (\bibinfo {year} {2016})},\ \Eprint {http://arxiv.org/abs/1606.07631}
  {arXiv:1606.07631 [astro-ph.CO]} \BibitemShut {NoStop}%
\bibitem [{\citenamefont {Young}\ and\ \citenamefont
  {Byrnes}(2013)}]{Young:2013oia}%
  \BibitemOpen
  \bibfield  {author} {\bibinfo {author} {\bibfnamefont {S.}~\bibnamefont
  {Young}}\ and\ \bibinfo {author} {\bibfnamefont {C.~T.}\ \bibnamefont
  {Byrnes}},\ }\href {\doibase 10.1088/1475-7516/2013/08/052} {\bibfield
  {journal} {\bibinfo  {journal} {JCAP}\ }\textbf {\bibinfo {volume} {08}},\
  \bibinfo {pages} {052} (\bibinfo {year} {2013})},\ \Eprint
  {http://arxiv.org/abs/1307.4995} {arXiv:1307.4995 [astro-ph.CO]} \BibitemShut
  {NoStop}%
\bibitem [{\citenamefont {Gow}\ \emph {et~al.}(2021)\citenamefont {Gow},
  \citenamefont {Byrnes}, \citenamefont {Cole},\ and\ \citenamefont
  {Young}}]{Gow:2020bzo}%
  \BibitemOpen
  \bibfield  {author} {\bibinfo {author} {\bibfnamefont {A.~D.}\ \bibnamefont
  {Gow}}, \bibinfo {author} {\bibfnamefont {C.~T.}\ \bibnamefont {Byrnes}},
  \bibinfo {author} {\bibfnamefont {P.~S.}\ \bibnamefont {Cole}}, \ and\
  \bibinfo {author} {\bibfnamefont {S.}~\bibnamefont {Young}},\ }\href
  {\doibase 10.1088/1475-7516/2021/02/002} {\bibfield  {journal} {\bibinfo
  {journal} {JCAP}\ }\textbf {\bibinfo {volume} {02}},\ \bibinfo {pages} {002}
  (\bibinfo {year} {2021})},\ \Eprint {http://arxiv.org/abs/2008.03289}
  {arXiv:2008.03289 [astro-ph.CO]} \BibitemShut {NoStop}%
\bibitem [{\citenamefont {Wang}\ \emph {et~al.}(2018)\citenamefont {Wang},
  \citenamefont {Wang}, \citenamefont {Huang},\ and\ \citenamefont
  {Li}}]{Wang:2016ana}%
  \BibitemOpen
  \bibfield  {author} {\bibinfo {author} {\bibfnamefont {S.}~\bibnamefont
  {Wang}}, \bibinfo {author} {\bibfnamefont {Y.-F.}\ \bibnamefont {Wang}},
  \bibinfo {author} {\bibfnamefont {Q.-G.}\ \bibnamefont {Huang}}, \ and\
  \bibinfo {author} {\bibfnamefont {T.~G.~F.}\ \bibnamefont {Li}},\ }\href
  {\doibase 10.1103/PhysRevLett.120.191102} {\bibfield  {journal} {\bibinfo
  {journal} {Phys. Rev. Lett.}\ }\textbf {\bibinfo {volume} {120}},\ \bibinfo
  {pages} {191102} (\bibinfo {year} {2018})},\ \Eprint
  {http://arxiv.org/abs/1610.08725} {arXiv:1610.08725 [astro-ph.CO]}
  \BibitemShut {NoStop}%
\bibitem [{\citenamefont {Wang}\ \emph {et~al.}(2021)\citenamefont {Wang},
  \citenamefont {Kohri},\ and\ \citenamefont {Vardanyan}}]{Wang:2021djr}%
  \BibitemOpen
  \bibfield  {author} {\bibinfo {author} {\bibfnamefont {S.}~\bibnamefont
  {Wang}}, \bibinfo {author} {\bibfnamefont {K.}~\bibnamefont {Kohri}}, \ and\
  \bibinfo {author} {\bibfnamefont {V.}~\bibnamefont {Vardanyan}},\ }\href@noop
  {} {\  (\bibinfo {year} {2021})},\ \Eprint {http://arxiv.org/abs/2107.01935}
  {arXiv:2107.01935 [gr-qc]} \BibitemShut {NoStop}%
\bibitem [{\citenamefont {Mollerach}\ \emph {et~al.}(2004)\citenamefont
  {Mollerach}, \citenamefont {Harari},\ and\ \citenamefont
  {Matarrese}}]{Mollerach:2003nq}%
  \BibitemOpen
  \bibfield  {author} {\bibinfo {author} {\bibfnamefont {S.}~\bibnamefont
  {Mollerach}}, \bibinfo {author} {\bibfnamefont {D.}~\bibnamefont {Harari}}, \
  and\ \bibinfo {author} {\bibfnamefont {S.}~\bibnamefont {Matarrese}},\ }\href
  {\doibase 10.1103/PhysRevD.69.063002} {\bibfield  {journal} {\bibinfo
  {journal} {Phys. Rev. D}\ }\textbf {\bibinfo {volume} {69}},\ \bibinfo
  {pages} {063002} (\bibinfo {year} {2004})},\ \Eprint
  {http://arxiv.org/abs/astro-ph/0310711} {arXiv:astro-ph/0310711} \BibitemShut
  {NoStop}%
\bibitem [{\citenamefont {Ananda}\ \emph {et~al.}(2007)\citenamefont {Ananda},
  \citenamefont {Clarkson},\ and\ \citenamefont {Wands}}]{Ananda:2006af}%
  \BibitemOpen
  \bibfield  {author} {\bibinfo {author} {\bibfnamefont {K.~N.}\ \bibnamefont
  {Ananda}}, \bibinfo {author} {\bibfnamefont {C.}~\bibnamefont {Clarkson}}, \
  and\ \bibinfo {author} {\bibfnamefont {D.}~\bibnamefont {Wands}},\ }\href
  {\doibase 10.1103/PhysRevD.75.123518} {\bibfield  {journal} {\bibinfo
  {journal} {Phys. Rev. D}\ }\textbf {\bibinfo {volume} {75}},\ \bibinfo
  {pages} {123518} (\bibinfo {year} {2007})},\ \Eprint
  {http://arxiv.org/abs/gr-qc/0612013} {arXiv:gr-qc/0612013} \BibitemShut
  {NoStop}%
\bibitem [{\citenamefont {Baumann}\ \emph {et~al.}(2007)\citenamefont
  {Baumann}, \citenamefont {Steinhardt}, \citenamefont {Takahashi},\ and\
  \citenamefont {Ichiki}}]{Baumann:2007zm}%
  \BibitemOpen
  \bibfield  {author} {\bibinfo {author} {\bibfnamefont {D.}~\bibnamefont
  {Baumann}}, \bibinfo {author} {\bibfnamefont {P.~J.}\ \bibnamefont
  {Steinhardt}}, \bibinfo {author} {\bibfnamefont {K.}~\bibnamefont
  {Takahashi}}, \ and\ \bibinfo {author} {\bibfnamefont {K.}~\bibnamefont
  {Ichiki}},\ }\href {\doibase 10.1103/PhysRevD.76.084019} {\bibfield
  {journal} {\bibinfo  {journal} {Phys. Rev. D}\ }\textbf {\bibinfo {volume}
  {76}},\ \bibinfo {pages} {084019} (\bibinfo {year} {2007})},\ \Eprint
  {http://arxiv.org/abs/hep-th/0703290} {arXiv:hep-th/0703290} \BibitemShut
  {NoStop}%
\bibitem [{\citenamefont {Kohri}\ and\ \citenamefont
  {Terada}(2018)}]{Kohri:2018awv}%
  \BibitemOpen
  \bibfield  {author} {\bibinfo {author} {\bibfnamefont {K.}~\bibnamefont
  {Kohri}}\ and\ \bibinfo {author} {\bibfnamefont {T.}~\bibnamefont {Terada}},\
  }\href {\doibase 10.1103/PhysRevD.97.123532} {\bibfield  {journal} {\bibinfo
  {journal} {Phys. Rev. D}\ }\textbf {\bibinfo {volume} {97}},\ \bibinfo
  {pages} {123532} (\bibinfo {year} {2018})},\ \Eprint
  {http://arxiv.org/abs/1804.08577} {arXiv:1804.08577 [gr-qc]} \BibitemShut
  {NoStop}%
\bibitem [{\citenamefont {Dom\`enech}(2021)}]{Domenech:2021ztg}%
  \BibitemOpen
  \bibfield  {author} {\bibinfo {author} {\bibfnamefont {G.}~\bibnamefont
  {Dom\`enech}},\ }\href {\doibase 10.3390/universe7110398} {\bibfield
  {journal} {\bibinfo  {journal} {Universe}\ }\textbf {\bibinfo {volume} {7}},\
  \bibinfo {pages} {398} (\bibinfo {year} {2021})},\ \Eprint
  {http://arxiv.org/abs/2109.01398} {arXiv:2109.01398 [gr-qc]} \BibitemShut
  {NoStop}%
\bibitem [{\citenamefont {Saito}\ and\ \citenamefont
  {Yokoyama}(2009)}]{Saito:2008jc}%
  \BibitemOpen
  \bibfield  {author} {\bibinfo {author} {\bibfnamefont {R.}~\bibnamefont
  {Saito}}\ and\ \bibinfo {author} {\bibfnamefont {J.}~\bibnamefont
  {Yokoyama}},\ }\href {\doibase 10.1103/PhysRevLett.102.161101} {\bibfield
  {journal} {\bibinfo  {journal} {Phys. Rev. Lett.}\ }\textbf {\bibinfo
  {volume} {102}},\ \bibinfo {pages} {161101} (\bibinfo {year} {2009})},\
  \bibinfo {note} {[Erratum: Phys.Rev.Lett. 107, 069901 (2011)]},\ \Eprint
  {http://arxiv.org/abs/0812.4339} {arXiv:0812.4339 [astro-ph]} \BibitemShut
  {NoStop}%
\bibitem [{\citenamefont {Wang}\ \emph {et~al.}(2019)\citenamefont {Wang},
  \citenamefont {Terada},\ and\ \citenamefont {Kohri}}]{Wang:2019kaf}%
  \BibitemOpen
  \bibfield  {author} {\bibinfo {author} {\bibfnamefont {S.}~\bibnamefont
  {Wang}}, \bibinfo {author} {\bibfnamefont {T.}~\bibnamefont {Terada}}, \ and\
  \bibinfo {author} {\bibfnamefont {K.}~\bibnamefont {Kohri}},\ }\href
  {\doibase 10.1103/PhysRevD.99.103531} {\bibfield  {journal} {\bibinfo
  {journal} {Phys. Rev. D}\ }\textbf {\bibinfo {volume} {99}},\ \bibinfo
  {pages} {103531} (\bibinfo {year} {2019})},\ \bibinfo {note} {[Erratum:
  Phys.Rev.D 101, 069901 (2020)]},\ \Eprint {http://arxiv.org/abs/1903.05924}
  {arXiv:1903.05924 [astro-ph.CO]} \BibitemShut {NoStop}%
\bibitem [{\citenamefont {Inomata}\ and\ \citenamefont
  {Nakama}(2019)}]{Inomata:2018epa}%
  \BibitemOpen
  \bibfield  {author} {\bibinfo {author} {\bibfnamefont {K.}~\bibnamefont
  {Inomata}}\ and\ \bibinfo {author} {\bibfnamefont {T.}~\bibnamefont
  {Nakama}},\ }\href {\doibase 10.1103/PhysRevD.99.043511} {\bibfield
  {journal} {\bibinfo  {journal} {Phys. Rev. D}\ }\textbf {\bibinfo {volume}
  {99}},\ \bibinfo {pages} {043511} (\bibinfo {year} {2019})},\ \Eprint
  {http://arxiv.org/abs/1812.00674} {arXiv:1812.00674 [astro-ph.CO]}
  \BibitemShut {NoStop}%
\bibitem [{\citenamefont {Byrnes}\ \emph {et~al.}(2019)\citenamefont {Byrnes},
  \citenamefont {Cole},\ and\ \citenamefont {Patil}}]{Byrnes:2018txb}%
  \BibitemOpen
  \bibfield  {author} {\bibinfo {author} {\bibfnamefont {C.~T.}\ \bibnamefont
  {Byrnes}}, \bibinfo {author} {\bibfnamefont {P.~S.}\ \bibnamefont {Cole}}, \
  and\ \bibinfo {author} {\bibfnamefont {S.~P.}\ \bibnamefont {Patil}},\ }\href
  {\doibase 10.1088/1475-7516/2019/06/028} {\bibfield  {journal} {\bibinfo
  {journal} {JCAP}\ }\textbf {\bibinfo {volume} {06}},\ \bibinfo {pages} {028}
  (\bibinfo {year} {2019})},\ \Eprint {http://arxiv.org/abs/1811.11158}
  {arXiv:1811.11158 [astro-ph.CO]} \BibitemShut {NoStop}%
\bibitem [{\citenamefont {Garcia-Bellido}\ \emph {et~al.}(2017)\citenamefont
  {Garcia-Bellido}, \citenamefont {Peloso},\ and\ \citenamefont
  {Unal}}]{Garcia-Bellido:2017aan}%
  \BibitemOpen
  \bibfield  {author} {\bibinfo {author} {\bibfnamefont {J.}~\bibnamefont
  {Garcia-Bellido}}, \bibinfo {author} {\bibfnamefont {M.}~\bibnamefont
  {Peloso}}, \ and\ \bibinfo {author} {\bibfnamefont {C.}~\bibnamefont
  {Unal}},\ }\href {\doibase 10.1088/1475-7516/2017/09/013} {\bibfield
  {journal} {\bibinfo  {journal} {JCAP}\ }\textbf {\bibinfo {volume} {09}},\
  \bibinfo {pages} {013} (\bibinfo {year} {2017})},\ \Eprint
  {http://arxiv.org/abs/1707.02441} {arXiv:1707.02441 [astro-ph.CO]}
  \BibitemShut {NoStop}%
\bibitem [{\citenamefont {Nakama}\ \emph {et~al.}(2017)\citenamefont {Nakama},
  \citenamefont {Silk},\ and\ \citenamefont {Kamionkowski}}]{Nakama:2016gzw}%
  \BibitemOpen
  \bibfield  {author} {\bibinfo {author} {\bibfnamefont {T.}~\bibnamefont
  {Nakama}}, \bibinfo {author} {\bibfnamefont {J.}~\bibnamefont {Silk}}, \ and\
  \bibinfo {author} {\bibfnamefont {M.}~\bibnamefont {Kamionkowski}},\ }\href
  {\doibase 10.1103/PhysRevD.95.043511} {\bibfield  {journal} {\bibinfo
  {journal} {Phys. Rev. D}\ }\textbf {\bibinfo {volume} {95}},\ \bibinfo
  {pages} {043511} (\bibinfo {year} {2017})},\ \Eprint
  {http://arxiv.org/abs/1612.06264} {arXiv:1612.06264 [astro-ph.CO]}
  \BibitemShut {NoStop}%
\bibitem [{\citenamefont {Bugaev}\ and\ \citenamefont
  {Klimai}(2011)}]{Bugaev:2010bb}%
  \BibitemOpen
  \bibfield  {author} {\bibinfo {author} {\bibfnamefont {E.}~\bibnamefont
  {Bugaev}}\ and\ \bibinfo {author} {\bibfnamefont {P.}~\bibnamefont
  {Klimai}},\ }\href {\doibase 10.1103/PhysRevD.83.083521} {\bibfield
  {journal} {\bibinfo  {journal} {Phys. Rev. D}\ }\textbf {\bibinfo {volume}
  {83}},\ \bibinfo {pages} {083521} (\bibinfo {year} {2011})},\ \Eprint
  {http://arxiv.org/abs/1012.4697} {arXiv:1012.4697 [astro-ph.CO]} \BibitemShut
  {NoStop}%
\bibitem [{\citenamefont {Saito}\ and\ \citenamefont
  {Yokoyama}(2010)}]{Saito:2009jt}%
  \BibitemOpen
  \bibfield  {author} {\bibinfo {author} {\bibfnamefont {R.}~\bibnamefont
  {Saito}}\ and\ \bibinfo {author} {\bibfnamefont {J.}~\bibnamefont
  {Yokoyama}},\ }\href {\doibase 10.1143/PTP.126.351} {\bibfield  {journal}
  {\bibinfo  {journal} {Prog. Theor. Phys.}\ }\textbf {\bibinfo {volume}
  {123}},\ \bibinfo {pages} {867} (\bibinfo {year} {2010})},\ \bibinfo {note}
  {[Erratum: Prog.Theor.Phys. 126, 351--352 (2011)]},\ \Eprint
  {http://arxiv.org/abs/0912.5317} {arXiv:0912.5317 [astro-ph.CO]} \BibitemShut
  {NoStop}%
\bibitem [{\citenamefont {Bugaev}\ and\ \citenamefont
  {Klimai}(2010)}]{Bugaev:2009zh}%
  \BibitemOpen
  \bibfield  {author} {\bibinfo {author} {\bibfnamefont {E.}~\bibnamefont
  {Bugaev}}\ and\ \bibinfo {author} {\bibfnamefont {P.}~\bibnamefont
  {Klimai}},\ }\href {\doibase 10.1103/PhysRevD.81.023517} {\bibfield
  {journal} {\bibinfo  {journal} {Phys. Rev. D}\ }\textbf {\bibinfo {volume}
  {81}},\ \bibinfo {pages} {023517} (\bibinfo {year} {2010})},\ \Eprint
  {http://arxiv.org/abs/0908.0664} {arXiv:0908.0664 [astro-ph.CO]} \BibitemShut
  {NoStop}%
\bibitem [{\citenamefont {Inomata}\ \emph {et~al.}(2020)\citenamefont
  {Inomata}, \citenamefont {Kawasaki}, \citenamefont {Mukaida}, \citenamefont
  {Terada},\ and\ \citenamefont {Yanagida}}]{Inomata:2020lmk}%
  \BibitemOpen
  \bibfield  {author} {\bibinfo {author} {\bibfnamefont {K.}~\bibnamefont
  {Inomata}}, \bibinfo {author} {\bibfnamefont {M.}~\bibnamefont {Kawasaki}},
  \bibinfo {author} {\bibfnamefont {K.}~\bibnamefont {Mukaida}}, \bibinfo
  {author} {\bibfnamefont {T.}~\bibnamefont {Terada}}, \ and\ \bibinfo {author}
  {\bibfnamefont {T.~T.}\ \bibnamefont {Yanagida}},\ }\href {\doibase
  10.1103/PhysRevD.101.123533} {\bibfield  {journal} {\bibinfo  {journal}
  {Phys. Rev. D}\ }\textbf {\bibinfo {volume} {101}},\ \bibinfo {pages}
  {123533} (\bibinfo {year} {2020})},\ \Eprint
  {http://arxiv.org/abs/2003.10455} {arXiv:2003.10455 [astro-ph.CO]}
  \BibitemShut {NoStop}%
\bibitem [{\citenamefont {Ballesteros}\ \emph {et~al.}(2020)\citenamefont
  {Ballesteros}, \citenamefont {Rey}, \citenamefont {Taoso},\ and\
  \citenamefont {Urbano}}]{Ballesteros:2020qam}%
  \BibitemOpen
  \bibfield  {author} {\bibinfo {author} {\bibfnamefont {G.}~\bibnamefont
  {Ballesteros}}, \bibinfo {author} {\bibfnamefont {J.}~\bibnamefont {Rey}},
  \bibinfo {author} {\bibfnamefont {M.}~\bibnamefont {Taoso}}, \ and\ \bibinfo
  {author} {\bibfnamefont {A.}~\bibnamefont {Urbano}},\ }\href {\doibase
  10.1088/1475-7516/2020/07/025} {\bibfield  {journal} {\bibinfo  {journal}
  {JCAP}\ }\textbf {\bibinfo {volume} {07}},\ \bibinfo {pages} {025} (\bibinfo
  {year} {2020})},\ \Eprint {http://arxiv.org/abs/2001.08220} {arXiv:2001.08220
  [astro-ph.CO]} \BibitemShut {NoStop}%
\bibitem [{\citenamefont {Lin}\ \emph {et~al.}(2020)\citenamefont {Lin},
  \citenamefont {Gao}, \citenamefont {Gong}, \citenamefont {Lu}, \citenamefont
  {Zhang},\ and\ \citenamefont {Zhang}}]{Lin:2020goi}%
  \BibitemOpen
  \bibfield  {author} {\bibinfo {author} {\bibfnamefont {J.}~\bibnamefont
  {Lin}}, \bibinfo {author} {\bibfnamefont {Q.}~\bibnamefont {Gao}}, \bibinfo
  {author} {\bibfnamefont {Y.}~\bibnamefont {Gong}}, \bibinfo {author}
  {\bibfnamefont {Y.}~\bibnamefont {Lu}}, \bibinfo {author} {\bibfnamefont
  {C.}~\bibnamefont {Zhang}}, \ and\ \bibinfo {author} {\bibfnamefont
  {F.}~\bibnamefont {Zhang}},\ }\href {\doibase 10.1103/PhysRevD.101.103515}
  {\bibfield  {journal} {\bibinfo  {journal} {Phys. Rev. D}\ }\textbf {\bibinfo
  {volume} {101}},\ \bibinfo {pages} {103515} (\bibinfo {year} {2020})},\
  \Eprint {http://arxiv.org/abs/2001.05909} {arXiv:2001.05909 [gr-qc]}
  \BibitemShut {NoStop}%
\bibitem [{\citenamefont {Chen}\ \emph {et~al.}(2020)\citenamefont {Chen},
  \citenamefont {Yuan},\ and\ \citenamefont {Huang}}]{Chen:2019xse}%
  \BibitemOpen
  \bibfield  {author} {\bibinfo {author} {\bibfnamefont {Z.-C.}\ \bibnamefont
  {Chen}}, \bibinfo {author} {\bibfnamefont {C.}~\bibnamefont {Yuan}}, \ and\
  \bibinfo {author} {\bibfnamefont {Q.-G.}\ \bibnamefont {Huang}},\ }\href
  {\doibase 10.1103/PhysRevLett.124.251101} {\bibfield  {journal} {\bibinfo
  {journal} {Phys. Rev. Lett.}\ }\textbf {\bibinfo {volume} {124}},\ \bibinfo
  {pages} {251101} (\bibinfo {year} {2020})},\ \Eprint
  {http://arxiv.org/abs/1910.12239} {arXiv:1910.12239 [astro-ph.CO]}
  \BibitemShut {NoStop}%
\bibitem [{\citenamefont {Cai}\ \emph {et~al.}(2019{\natexlab{a}})\citenamefont
  {Cai}, \citenamefont {Pi}, \citenamefont {Wang},\ and\ \citenamefont
  {Yang}}]{Cai:2019elf}%
  \BibitemOpen
  \bibfield  {author} {\bibinfo {author} {\bibfnamefont {R.-G.}\ \bibnamefont
  {Cai}}, \bibinfo {author} {\bibfnamefont {S.}~\bibnamefont {Pi}}, \bibinfo
  {author} {\bibfnamefont {S.-J.}\ \bibnamefont {Wang}}, \ and\ \bibinfo
  {author} {\bibfnamefont {X.-Y.}\ \bibnamefont {Yang}},\ }\href {\doibase
  10.1088/1475-7516/2019/10/059} {\bibfield  {journal} {\bibinfo  {journal}
  {JCAP}\ }\textbf {\bibinfo {volume} {10}},\ \bibinfo {pages} {059} (\bibinfo
  {year} {2019}{\natexlab{a}})},\ \Eprint {http://arxiv.org/abs/1907.06372}
  {arXiv:1907.06372 [astro-ph.CO]} \BibitemShut {NoStop}%
\bibitem [{\citenamefont {Cai}\ \emph {et~al.}(2019{\natexlab{b}})\citenamefont
  {Cai}, \citenamefont {Chen}, \citenamefont {Tong}, \citenamefont {Wang},\
  and\ \citenamefont {Yan}}]{Cai:2019jah}%
  \BibitemOpen
  \bibfield  {author} {\bibinfo {author} {\bibfnamefont {Y.-F.}\ \bibnamefont
  {Cai}}, \bibinfo {author} {\bibfnamefont {C.}~\bibnamefont {Chen}}, \bibinfo
  {author} {\bibfnamefont {X.}~\bibnamefont {Tong}}, \bibinfo {author}
  {\bibfnamefont {D.-G.}\ \bibnamefont {Wang}}, \ and\ \bibinfo {author}
  {\bibfnamefont {S.-F.}\ \bibnamefont {Yan}},\ }\href {\doibase
  10.1103/PhysRevD.100.043518} {\bibfield  {journal} {\bibinfo  {journal}
  {Phys. Rev. D}\ }\textbf {\bibinfo {volume} {100}},\ \bibinfo {pages}
  {043518} (\bibinfo {year} {2019}{\natexlab{b}})},\ \Eprint
  {http://arxiv.org/abs/1902.08187} {arXiv:1902.08187 [astro-ph.CO]}
  \BibitemShut {NoStop}%
\bibitem [{\citenamefont {Ando}\ \emph {et~al.}(2018)\citenamefont {Ando},
  \citenamefont {Inomata},\ and\ \citenamefont {Kawasaki}}]{Ando:2018qdb}%
  \BibitemOpen
  \bibfield  {author} {\bibinfo {author} {\bibfnamefont {K.}~\bibnamefont
  {Ando}}, \bibinfo {author} {\bibfnamefont {K.}~\bibnamefont {Inomata}}, \
  and\ \bibinfo {author} {\bibfnamefont {M.}~\bibnamefont {Kawasaki}},\ }\href
  {\doibase 10.1103/PhysRevD.97.103528} {\bibfield  {journal} {\bibinfo
  {journal} {Phys. Rev. D}\ }\textbf {\bibinfo {volume} {97}},\ \bibinfo
  {pages} {103528} (\bibinfo {year} {2018})},\ \Eprint
  {http://arxiv.org/abs/1802.06393} {arXiv:1802.06393 [astro-ph.CO]}
  \BibitemShut {NoStop}%
\bibitem [{\citenamefont {Di}\ and\ \citenamefont {Gong}(2018)}]{Di:2017ndc}%
  \BibitemOpen
  \bibfield  {author} {\bibinfo {author} {\bibfnamefont {H.}~\bibnamefont
  {Di}}\ and\ \bibinfo {author} {\bibfnamefont {Y.}~\bibnamefont {Gong}},\
  }\href {\doibase 10.1088/1475-7516/2018/07/007} {\bibfield  {journal}
  {\bibinfo  {journal} {JCAP}\ }\textbf {\bibinfo {volume} {07}},\ \bibinfo
  {pages} {007} (\bibinfo {year} {2018})},\ \Eprint
  {http://arxiv.org/abs/1707.09578} {arXiv:1707.09578 [astro-ph.CO]}
  \BibitemShut {NoStop}%
\bibitem [{\citenamefont {Chang}\ \emph
  {et~al.}(2022{\natexlab{a}})\citenamefont {Chang}, \citenamefont {Zhang},\
  and\ \citenamefont {Zhou}}]{Chang:2022dhh}%
  \BibitemOpen
  \bibfield  {author} {\bibinfo {author} {\bibfnamefont {Z.}~\bibnamefont
  {Chang}}, \bibinfo {author} {\bibfnamefont {X.}~\bibnamefont {Zhang}}, \ and\
  \bibinfo {author} {\bibfnamefont {J.-Z.}\ \bibnamefont {Zhou}},\ }\href@noop
  {} {\  (\bibinfo {year} {2022}{\natexlab{a}})},\ \Eprint
  {http://arxiv.org/abs/2207.01231} {arXiv:2207.01231 [astro-ph.CO]}
  \BibitemShut {NoStop}%
\bibitem [{\citenamefont {Chang}\ \emph
  {et~al.}(2022{\natexlab{b}})\citenamefont {Chang}, \citenamefont {Zhang},\
  and\ \citenamefont {Zhou}}]{Chang:2022vlv}%
  \BibitemOpen
  \bibfield  {author} {\bibinfo {author} {\bibfnamefont {Z.}~\bibnamefont
  {Chang}}, \bibinfo {author} {\bibfnamefont {X.}~\bibnamefont {Zhang}}, \ and\
  \bibinfo {author} {\bibfnamefont {J.-Z.}\ \bibnamefont {Zhou}},\ }\href@noop
  {} {\  (\bibinfo {year} {2022}{\natexlab{b}})},\ \Eprint
  {http://arxiv.org/abs/2209.07693} {arXiv:2209.07693 [astro-ph.CO]}
  \BibitemShut {NoStop}%
\bibitem [{\citenamefont {Orlofsky}\ \emph {et~al.}(2017)\citenamefont
  {Orlofsky}, \citenamefont {Pierce},\ and\ \citenamefont
  {Wells}}]{Orlofsky:2016vbd}%
  \BibitemOpen
  \bibfield  {author} {\bibinfo {author} {\bibfnamefont {N.}~\bibnamefont
  {Orlofsky}}, \bibinfo {author} {\bibfnamefont {A.}~\bibnamefont {Pierce}}, \
  and\ \bibinfo {author} {\bibfnamefont {J.~D.}\ \bibnamefont {Wells}},\ }\href
  {\doibase 10.1103/PhysRevD.95.063518} {\bibfield  {journal} {\bibinfo
  {journal} {Phys. Rev. D}\ }\textbf {\bibinfo {volume} {95}},\ \bibinfo
  {pages} {063518} (\bibinfo {year} {2017})},\ \Eprint
  {http://arxiv.org/abs/1612.05279} {arXiv:1612.05279 [astro-ph.CO]}
  \BibitemShut {NoStop}%
\bibitem [{\citenamefont {Gao}(2021)}]{Gao:2021vxb}%
  \BibitemOpen
  \bibfield  {author} {\bibinfo {author} {\bibfnamefont {Q.}~\bibnamefont
  {Gao}},\ }\href {\doibase 10.1007/s11433-021-1708-9} {\bibfield  {journal}
  {\bibinfo  {journal} {Sci. China Phys. Mech. Astron.}\ }\textbf {\bibinfo
  {volume} {64}},\ \bibinfo {pages} {280411} (\bibinfo {year} {2021})},\
  \Eprint {http://arxiv.org/abs/2102.07369} {arXiv:2102.07369 [gr-qc]}
  \BibitemShut {NoStop}%
\bibitem [{\citenamefont {Nakama}\ and\ \citenamefont
  {Suyama}(2015)}]{Nakama:2015nea}%
  \BibitemOpen
  \bibfield  {author} {\bibinfo {author} {\bibfnamefont {T.}~\bibnamefont
  {Nakama}}\ and\ \bibinfo {author} {\bibfnamefont {T.}~\bibnamefont
  {Suyama}},\ }\href {\doibase 10.1103/PhysRevD.92.121304} {\bibfield
  {journal} {\bibinfo  {journal} {Phys. Rev. D}\ }\textbf {\bibinfo {volume}
  {92}},\ \bibinfo {pages} {121304} (\bibinfo {year} {2015})},\ \Eprint
  {http://arxiv.org/abs/1506.05228} {arXiv:1506.05228 [gr-qc]} \BibitemShut
  {NoStop}%
\bibitem [{\citenamefont {Nakama}\ and\ \citenamefont
  {Suyama}(2016)}]{Nakama:2016enz}%
  \BibitemOpen
  \bibfield  {author} {\bibinfo {author} {\bibfnamefont {T.}~\bibnamefont
  {Nakama}}\ and\ \bibinfo {author} {\bibfnamefont {T.}~\bibnamefont
  {Suyama}},\ }\href {\doibase 10.1103/PhysRevD.94.043507} {\bibfield
  {journal} {\bibinfo  {journal} {Phys. Rev. D}\ }\textbf {\bibinfo {volume}
  {94}},\ \bibinfo {pages} {043507} (\bibinfo {year} {2016})},\ \Eprint
  {http://arxiv.org/abs/1605.04482} {arXiv:1605.04482 [gr-qc]} \BibitemShut
  {NoStop}%
\bibitem [{\citenamefont {Changa}\ \emph {et~al.}(2022)\citenamefont {Changa},
  \citenamefont {Tinga}, \citenamefont {Zhang},\ and\ \citenamefont
  {Zhou}}]{Changa:2022trj}%
  \BibitemOpen
  \bibfield  {author} {\bibinfo {author} {\bibfnamefont {Z.}~\bibnamefont
  {Changa}}, \bibinfo {author} {\bibfnamefont {K.-Y.}\ \bibnamefont {Tinga}},
  \bibinfo {author} {\bibfnamefont {X.}~\bibnamefont {Zhang}}, \ and\ \bibinfo
  {author} {\bibfnamefont {J.-Z.}\ \bibnamefont {Zhou}},\ }\href@noop {} {\
  (\bibinfo {year} {2022})},\ \Eprint {http://arxiv.org/abs/2211.11948}
  {arXiv:2211.11948 [astro-ph.CO]} \BibitemShut {NoStop}%
\bibitem [{\citenamefont {Inomata}(2021)}]{Inomata:2020cck}%
  \BibitemOpen
  \bibfield  {author} {\bibinfo {author} {\bibfnamefont {K.}~\bibnamefont
  {Inomata}},\ }\href {\doibase 10.1088/1475-7516/2021/03/013} {\bibfield
  {journal} {\bibinfo  {journal} {JCAP}\ }\textbf {\bibinfo {volume} {03}},\
  \bibinfo {pages} {013} (\bibinfo {year} {2021})},\ \Eprint
  {http://arxiv.org/abs/2008.12300} {arXiv:2008.12300 [gr-qc]} \BibitemShut
  {NoStop}%
\bibitem [{\citenamefont {Cai}\ \emph {et~al.}(2019{\natexlab{c}})\citenamefont
  {Cai}, \citenamefont {Pi},\ and\ \citenamefont {Sasaki}}]{Cai:2018dig}%
  \BibitemOpen
  \bibfield  {author} {\bibinfo {author} {\bibfnamefont {R.-g.}\ \bibnamefont
  {Cai}}, \bibinfo {author} {\bibfnamefont {S.}~\bibnamefont {Pi}}, \ and\
  \bibinfo {author} {\bibfnamefont {M.}~\bibnamefont {Sasaki}},\ }\href
  {\doibase 10.1103/PhysRevLett.122.201101} {\bibfield  {journal} {\bibinfo
  {journal} {Phys. Rev. Lett.}\ }\textbf {\bibinfo {volume} {122}},\ \bibinfo
  {pages} {201101} (\bibinfo {year} {2019}{\natexlab{c}})},\ \Eprint
  {http://arxiv.org/abs/1810.11000} {arXiv:1810.11000 [astro-ph.CO]}
  \BibitemShut {NoStop}%
\bibitem [{\citenamefont {Atal}\ and\ \citenamefont
  {Dom\`enech}(2021)}]{Atal:2021jyo}%
  \BibitemOpen
  \bibfield  {author} {\bibinfo {author} {\bibfnamefont {V.}~\bibnamefont
  {Atal}}\ and\ \bibinfo {author} {\bibfnamefont {G.}~\bibnamefont
  {Dom\`enech}},\ }\href {\doibase 10.1088/1475-7516/2021/06/001} {\bibfield
  {journal} {\bibinfo  {journal} {JCAP}\ }\textbf {\bibinfo {volume} {06}},\
  \bibinfo {pages} {001} (\bibinfo {year} {2021})},\ \Eprint
  {http://arxiv.org/abs/2103.01056} {arXiv:2103.01056 [astro-ph.CO]}
  \BibitemShut {NoStop}%
\bibitem [{\citenamefont {Zhang}\ \emph {et~al.}(2021)\citenamefont {Zhang},
  \citenamefont {Gong}, \citenamefont {Lin}, \citenamefont {Lu},\ and\
  \citenamefont {Yi}}]{Zhang:2020uek}%
  \BibitemOpen
  \bibfield  {author} {\bibinfo {author} {\bibfnamefont {F.}~\bibnamefont
  {Zhang}}, \bibinfo {author} {\bibfnamefont {Y.}~\bibnamefont {Gong}},
  \bibinfo {author} {\bibfnamefont {J.}~\bibnamefont {Lin}}, \bibinfo {author}
  {\bibfnamefont {Y.}~\bibnamefont {Lu}}, \ and\ \bibinfo {author}
  {\bibfnamefont {Z.}~\bibnamefont {Yi}},\ }\href {\doibase
  10.1088/1475-7516/2021/04/045} {\bibfield  {journal} {\bibinfo  {journal}
  {JCAP}\ }\textbf {\bibinfo {volume} {04}},\ \bibinfo {pages} {045} (\bibinfo
  {year} {2021})},\ \Eprint {http://arxiv.org/abs/2012.06960} {arXiv:2012.06960
  [astro-ph.CO]} \BibitemShut {NoStop}%
\bibitem [{\citenamefont {Yuan}\ and\ \citenamefont
  {Huang}(2021)}]{Yuan:2020iwf}%
  \BibitemOpen
  \bibfield  {author} {\bibinfo {author} {\bibfnamefont {C.}~\bibnamefont
  {Yuan}}\ and\ \bibinfo {author} {\bibfnamefont {Q.-G.}\ \bibnamefont
  {Huang}},\ }\href {\doibase 10.1016/j.physletb.2021.136606} {\bibfield
  {journal} {\bibinfo  {journal} {Phys. Lett. B}\ }\textbf {\bibinfo {volume}
  {821}},\ \bibinfo {pages} {136606} (\bibinfo {year} {2021})},\ \Eprint
  {http://arxiv.org/abs/2007.10686} {arXiv:2007.10686 [astro-ph.CO]}
  \BibitemShut {NoStop}%
\bibitem [{\citenamefont {Davies}\ \emph {et~al.}(2022)\citenamefont {Davies},
  \citenamefont {Carrilho},\ and\ \citenamefont {Mulryne}}]{Davies:2021loj}%
  \BibitemOpen
  \bibfield  {author} {\bibinfo {author} {\bibfnamefont {M.~W.}\ \bibnamefont
  {Davies}}, \bibinfo {author} {\bibfnamefont {P.}~\bibnamefont {Carrilho}}, \
  and\ \bibinfo {author} {\bibfnamefont {D.~J.}\ \bibnamefont {Mulryne}},\
  }\href {\doibase 10.1088/1475-7516/2022/06/019} {\bibfield  {journal}
  {\bibinfo  {journal} {JCAP}\ }\textbf {\bibinfo {volume} {06}},\ \bibinfo
  {pages} {019} (\bibinfo {year} {2022})},\ \Eprint
  {http://arxiv.org/abs/2110.08189} {arXiv:2110.08189 [astro-ph.CO]}
  \BibitemShut {NoStop}%
\bibitem [{\citenamefont {Rezazadeh}\ \emph {et~al.}(2022)\citenamefont
  {Rezazadeh}, \citenamefont {Teimoori}, \citenamefont {Karimi},\ and\
  \citenamefont {Karami}}]{Rezazadeh:2021clf}%
  \BibitemOpen
  \bibfield  {author} {\bibinfo {author} {\bibfnamefont {K.}~\bibnamefont
  {Rezazadeh}}, \bibinfo {author} {\bibfnamefont {Z.}~\bibnamefont {Teimoori}},
  \bibinfo {author} {\bibfnamefont {S.}~\bibnamefont {Karimi}}, \ and\ \bibinfo
  {author} {\bibfnamefont {K.}~\bibnamefont {Karami}},\ }\href {\doibase
  10.1140/epjc/s10052-022-10735-w} {\bibfield  {journal} {\bibinfo  {journal}
  {Eur. Phys. J. C}\ }\textbf {\bibinfo {volume} {82}},\ \bibinfo {pages} {758}
  (\bibinfo {year} {2022})},\ \Eprint {http://arxiv.org/abs/2110.01482}
  {arXiv:2110.01482 [gr-qc]} \BibitemShut {NoStop}%
\bibitem [{\citenamefont {Kristiano}\ and\ \citenamefont
  {Yokoyama}(2022)}]{Kristiano:2021urj}%
  \BibitemOpen
  \bibfield  {author} {\bibinfo {author} {\bibfnamefont {J.}~\bibnamefont
  {Kristiano}}\ and\ \bibinfo {author} {\bibfnamefont {J.}~\bibnamefont
  {Yokoyama}},\ }\href {\doibase 10.1103/PhysRevLett.128.061301} {\bibfield
  {journal} {\bibinfo  {journal} {Phys. Rev. Lett.}\ }\textbf {\bibinfo
  {volume} {128}},\ \bibinfo {pages} {061301} (\bibinfo {year} {2022})},\
  \Eprint {http://arxiv.org/abs/2104.01953} {arXiv:2104.01953 [hep-th]}
  \BibitemShut {NoStop}%
\bibitem [{\citenamefont {Bartolo}\ \emph {et~al.}(2018)\citenamefont
  {Bartolo}, \citenamefont {Domcke}, \citenamefont {Figueroa}, \citenamefont
  {Garc\'\i{}a-Bellido}, \citenamefont {Peloso}, \citenamefont {Pieroni},
  \citenamefont {Ricciardone}, \citenamefont {Sakellariadou}, \citenamefont
  {Sorbo},\ and\ \citenamefont {Tasinato}}]{Bartolo:2018qqn}%
  \BibitemOpen
  \bibfield  {author} {\bibinfo {author} {\bibfnamefont {N.}~\bibnamefont
  {Bartolo}}, \bibinfo {author} {\bibfnamefont {V.}~\bibnamefont {Domcke}},
  \bibinfo {author} {\bibfnamefont {D.~G.}\ \bibnamefont {Figueroa}}, \bibinfo
  {author} {\bibfnamefont {J.}~\bibnamefont {Garc\'\i{}a-Bellido}}, \bibinfo
  {author} {\bibfnamefont {M.}~\bibnamefont {Peloso}}, \bibinfo {author}
  {\bibfnamefont {M.}~\bibnamefont {Pieroni}}, \bibinfo {author} {\bibfnamefont
  {A.}~\bibnamefont {Ricciardone}}, \bibinfo {author} {\bibfnamefont
  {M.}~\bibnamefont {Sakellariadou}}, \bibinfo {author} {\bibfnamefont
  {L.}~\bibnamefont {Sorbo}}, \ and\ \bibinfo {author} {\bibfnamefont
  {G.}~\bibnamefont {Tasinato}},\ }\href {\doibase
  10.1088/1475-7516/2018/11/034} {\bibfield  {journal} {\bibinfo  {journal}
  {JCAP}\ }\textbf {\bibinfo {volume} {11}},\ \bibinfo {pages} {034} (\bibinfo
  {year} {2018})},\ \Eprint {http://arxiv.org/abs/1806.02819} {arXiv:1806.02819
  [astro-ph.CO]} \BibitemShut {NoStop}%
\bibitem [{\citenamefont {Hwang}\ \emph {et~al.}(2017)\citenamefont {Hwang},
  \citenamefont {Jeong},\ and\ \citenamefont {Noh}}]{Hwang:2017oxa}%
  \BibitemOpen
  \bibfield  {author} {\bibinfo {author} {\bibfnamefont {J.-C.}\ \bibnamefont
  {Hwang}}, \bibinfo {author} {\bibfnamefont {D.}~\bibnamefont {Jeong}}, \ and\
  \bibinfo {author} {\bibfnamefont {H.}~\bibnamefont {Noh}},\ }\href {\doibase
  10.3847/1538-4357/aa74be} {\bibfield  {journal} {\bibinfo  {journal}
  {Astrophys. J.}\ }\textbf {\bibinfo {volume} {842}},\ \bibinfo {pages} {46}
  (\bibinfo {year} {2017})},\ \Eprint {http://arxiv.org/abs/1704.03500}
  {arXiv:1704.03500 [astro-ph.CO]} \BibitemShut {NoStop}%
\bibitem [{\citenamefont {Yuan}\ \emph {et~al.}(2020)\citenamefont {Yuan},
  \citenamefont {Chen},\ and\ \citenamefont {Huang}}]{Yuan:2019fwv}%
  \BibitemOpen
  \bibfield  {author} {\bibinfo {author} {\bibfnamefont {C.}~\bibnamefont
  {Yuan}}, \bibinfo {author} {\bibfnamefont {Z.-C.}\ \bibnamefont {Chen}}, \
  and\ \bibinfo {author} {\bibfnamefont {Q.-G.}\ \bibnamefont {Huang}},\ }\href
  {\doibase 10.1103/PhysRevD.101.063018} {\bibfield  {journal} {\bibinfo
  {journal} {Phys. Rev. D}\ }\textbf {\bibinfo {volume} {101}},\ \bibinfo
  {pages} {063018} (\bibinfo {year} {2020})},\ \Eprint
  {http://arxiv.org/abs/1912.00885} {arXiv:1912.00885 [astro-ph.CO]}
  \BibitemShut {NoStop}%
\bibitem [{\citenamefont {Inomata}\ and\ \citenamefont
  {Terada}(2020)}]{Inomata:2019yww}%
  \BibitemOpen
  \bibfield  {author} {\bibinfo {author} {\bibfnamefont {K.}~\bibnamefont
  {Inomata}}\ and\ \bibinfo {author} {\bibfnamefont {T.}~\bibnamefont
  {Terada}},\ }\href {\doibase 10.1103/PhysRevD.101.023523} {\bibfield
  {journal} {\bibinfo  {journal} {Phys. Rev. D}\ }\textbf {\bibinfo {volume}
  {101}},\ \bibinfo {pages} {023523} (\bibinfo {year} {2020})},\ \Eprint
  {http://arxiv.org/abs/1912.00785} {arXiv:1912.00785 [gr-qc]} \BibitemShut
  {NoStop}%
\bibitem [{\citenamefont {De~Luca}\ \emph
  {et~al.}(2020{\natexlab{b}})\citenamefont {De~Luca}, \citenamefont
  {Franciolini}, \citenamefont {Kehagias},\ and\ \citenamefont
  {Riotto}}]{DeLuca:2019ufz}%
  \BibitemOpen
  \bibfield  {author} {\bibinfo {author} {\bibfnamefont {V.}~\bibnamefont
  {De~Luca}}, \bibinfo {author} {\bibfnamefont {G.}~\bibnamefont
  {Franciolini}}, \bibinfo {author} {\bibfnamefont {A.}~\bibnamefont
  {Kehagias}}, \ and\ \bibinfo {author} {\bibfnamefont {A.}~\bibnamefont
  {Riotto}},\ }\href {\doibase 10.1088/1475-7516/2020/03/014} {\bibfield
  {journal} {\bibinfo  {journal} {JCAP}\ }\textbf {\bibinfo {volume} {03}},\
  \bibinfo {pages} {014} (\bibinfo {year} {2020}{\natexlab{b}})},\ \Eprint
  {http://arxiv.org/abs/1911.09689} {arXiv:1911.09689 [gr-qc]} \BibitemShut
  {NoStop}%
\bibitem [{\citenamefont {Dom\`enech}\ and\ \citenamefont
  {Sasaki}(2021)}]{Domenech:2020xin}%
  \BibitemOpen
  \bibfield  {author} {\bibinfo {author} {\bibfnamefont {G.}~\bibnamefont
  {Dom\`enech}}\ and\ \bibinfo {author} {\bibfnamefont {M.}~\bibnamefont
  {Sasaki}},\ }\href {\doibase 10.1103/PhysRevD.103.063531} {\bibfield
  {journal} {\bibinfo  {journal} {Phys. Rev. D}\ }\textbf {\bibinfo {volume}
  {103}},\ \bibinfo {pages} {063531} (\bibinfo {year} {2021})},\ \Eprint
  {http://arxiv.org/abs/2012.14016} {arXiv:2012.14016 [gr-qc]} \BibitemShut
  {NoStop}%
\bibitem [{\citenamefont {Chang}\ \emph {et~al.}(2021)\citenamefont {Chang},
  \citenamefont {Wang},\ and\ \citenamefont {Zhu}}]{Chang:2020tji}%
  \BibitemOpen
  \bibfield  {author} {\bibinfo {author} {\bibfnamefont {Z.}~\bibnamefont
  {Chang}}, \bibinfo {author} {\bibfnamefont {S.}~\bibnamefont {Wang}}, \ and\
  \bibinfo {author} {\bibfnamefont {Q.-H.}\ \bibnamefont {Zhu}},\ }\href
  {\doibase 10.1088/1674-1137/ac0c74} {\bibfield  {journal} {\bibinfo
  {journal} {Chin. Phys. C}\ }\textbf {\bibinfo {volume} {45}},\ \bibinfo
  {pages} {095101} (\bibinfo {year} {2021})},\ \Eprint
  {http://arxiv.org/abs/2009.11025} {arXiv:2009.11025 [astro-ph.CO]}
  \BibitemShut {NoStop}%
\bibitem [{\citenamefont {Ali}\ \emph {et~al.}(2021)\citenamefont {Ali},
  \citenamefont {Gong},\ and\ \citenamefont {Lu}}]{Ali:2020sfw}%
  \BibitemOpen
  \bibfield  {author} {\bibinfo {author} {\bibfnamefont {A.}~\bibnamefont
  {Ali}}, \bibinfo {author} {\bibfnamefont {Y.}~\bibnamefont {Gong}}, \ and\
  \bibinfo {author} {\bibfnamefont {Y.}~\bibnamefont {Lu}},\ }\href {\doibase
  10.1103/PhysRevD.103.043516} {\bibfield  {journal} {\bibinfo  {journal}
  {Phys. Rev. D}\ }\textbf {\bibinfo {volume} {103}},\ \bibinfo {pages}
  {043516} (\bibinfo {year} {2021})},\ \Eprint
  {http://arxiv.org/abs/2009.11081} {arXiv:2009.11081 [gr-qc]} \BibitemShut
  {NoStop}%
\bibitem [{\citenamefont {Lu}\ \emph {et~al.}(2020)\citenamefont {Lu},
  \citenamefont {Ali}, \citenamefont {Gong}, \citenamefont {Lin},\ and\
  \citenamefont {Zhang}}]{Lu:2020diy}%
  \BibitemOpen
  \bibfield  {author} {\bibinfo {author} {\bibfnamefont {Y.}~\bibnamefont
  {Lu}}, \bibinfo {author} {\bibfnamefont {A.}~\bibnamefont {Ali}}, \bibinfo
  {author} {\bibfnamefont {Y.}~\bibnamefont {Gong}}, \bibinfo {author}
  {\bibfnamefont {J.}~\bibnamefont {Lin}}, \ and\ \bibinfo {author}
  {\bibfnamefont {F.}~\bibnamefont {Zhang}},\ }\href {\doibase
  10.1103/PhysRevD.102.083503(2020)} {\bibfield  {journal} {\bibinfo  {journal}
  {Phys. Rev. D}\ }\textbf {\bibinfo {volume} {102}},\ \bibinfo {pages}
  {083503} (\bibinfo {year} {2020})},\ \Eprint
  {http://arxiv.org/abs/2006.03450} {arXiv:2006.03450 [gr-qc]} \BibitemShut
  {NoStop}%
\bibitem [{\citenamefont {Tomikawa}\ and\ \citenamefont
  {Kobayashi}(2020)}]{Tomikawa:2019tvi}%
  \BibitemOpen
  \bibfield  {author} {\bibinfo {author} {\bibfnamefont {K.}~\bibnamefont
  {Tomikawa}}\ and\ \bibinfo {author} {\bibfnamefont {T.}~\bibnamefont
  {Kobayashi}},\ }\href {\doibase 10.1103/PhysRevD.101.083529} {\bibfield
  {journal} {\bibinfo  {journal} {Phys. Rev. D}\ }\textbf {\bibinfo {volume}
  {101}},\ \bibinfo {pages} {083529} (\bibinfo {year} {2020})},\ \Eprint
  {http://arxiv.org/abs/1910.01880} {arXiv:1910.01880 [gr-qc]} \BibitemShut
  {NoStop}%
\bibitem [{\citenamefont {Gurian}\ \emph {et~al.}(2021)\citenamefont {Gurian},
  \citenamefont {Jeong}, \citenamefont {Hwang},\ and\ \citenamefont
  {Noh}}]{Gurian:2021rfv}%
  \BibitemOpen
  \bibfield  {author} {\bibinfo {author} {\bibfnamefont {J.}~\bibnamefont
  {Gurian}}, \bibinfo {author} {\bibfnamefont {D.}~\bibnamefont {Jeong}},
  \bibinfo {author} {\bibfnamefont {J.-c.}\ \bibnamefont {Hwang}}, \ and\
  \bibinfo {author} {\bibfnamefont {H.}~\bibnamefont {Noh}},\ }\href {\doibase
  10.1103/PhysRevD.104.083534} {\bibfield  {journal} {\bibinfo  {journal}
  {Phys. Rev. D}\ }\textbf {\bibinfo {volume} {104}},\ \bibinfo {pages}
  {083534} (\bibinfo {year} {2021})},\ \Eprint
  {http://arxiv.org/abs/2104.03330} {arXiv:2104.03330 [astro-ph.CO]}
  \BibitemShut {NoStop}%
\bibitem [{\citenamefont {Uggla}\ and\ \citenamefont
  {Wainwright}(2019)}]{Uggla:2018fiy}%
  \BibitemOpen
  \bibfield  {author} {\bibinfo {author} {\bibfnamefont {C.}~\bibnamefont
  {Uggla}}\ and\ \bibinfo {author} {\bibfnamefont {J.}~\bibnamefont
  {Wainwright}},\ }\href {\doibase 10.1088/1361-6382/aaf924} {\bibfield
  {journal} {\bibinfo  {journal} {Class. Quant. Grav.}\ }\textbf {\bibinfo
  {volume} {36}},\ \bibinfo {pages} {035004} (\bibinfo {year} {2019})},\
  \Eprint {http://arxiv.org/abs/1801.04300} {arXiv:1801.04300 [gr-qc]}
  \BibitemShut {NoStop}%
\bibitem [{\citenamefont {Dom\`enech}\ \emph {et~al.}(2022)\citenamefont
  {Dom\`enech}, \citenamefont {Passaglia},\ and\ \citenamefont
  {Renaux-Petel}}]{Domenech:2021and}%
  \BibitemOpen
  \bibfield  {author} {\bibinfo {author} {\bibfnamefont {G.}~\bibnamefont
  {Dom\`enech}}, \bibinfo {author} {\bibfnamefont {S.}~\bibnamefont
  {Passaglia}}, \ and\ \bibinfo {author} {\bibfnamefont {S.}~\bibnamefont
  {Renaux-Petel}},\ }\href {\doibase 10.1088/1475-7516/2022/03/023} {\bibfield
  {journal} {\bibinfo  {journal} {JCAP}\ }\textbf {\bibinfo {volume} {03}},\
  \bibinfo {pages} {023} (\bibinfo {year} {2022})},\ \Eprint
  {http://arxiv.org/abs/2112.10163} {arXiv:2112.10163 [astro-ph.CO]}
  \BibitemShut {NoStop}%
\bibitem [{\citenamefont {Zhang}\ \emph {et~al.}(2022)\citenamefont {Zhang},
  \citenamefont {Zhou},\ and\ \citenamefont {Chang}}]{Zhang:2022dgx}%
  \BibitemOpen
  \bibfield  {author} {\bibinfo {author} {\bibfnamefont {X.}~\bibnamefont
  {Zhang}}, \bibinfo {author} {\bibfnamefont {J.-Z.}\ \bibnamefont {Zhou}}, \
  and\ \bibinfo {author} {\bibfnamefont {Z.}~\bibnamefont {Chang}},\ }\href
  {\doibase 10.1140/epjc/s10052-022-10742-x} {\bibfield  {journal} {\bibinfo
  {journal} {Eur. Phys. J. C}\ }\textbf {\bibinfo {volume} {82}},\ \bibinfo
  {pages} {781} (\bibinfo {year} {2022})},\ \Eprint
  {http://arxiv.org/abs/2208.12948} {arXiv:2208.12948 [astro-ph.CO]}
  \BibitemShut {NoStop}%
\bibitem [{\citenamefont {Mangilli}\ \emph {et~al.}(2008)\citenamefont
  {Mangilli}, \citenamefont {Bartolo}, \citenamefont {Matarrese},\ and\
  \citenamefont {Riotto}}]{Mangilli:2008bw}%
  \BibitemOpen
  \bibfield  {author} {\bibinfo {author} {\bibfnamefont {A.}~\bibnamefont
  {Mangilli}}, \bibinfo {author} {\bibfnamefont {N.}~\bibnamefont {Bartolo}},
  \bibinfo {author} {\bibfnamefont {S.}~\bibnamefont {Matarrese}}, \ and\
  \bibinfo {author} {\bibfnamefont {A.}~\bibnamefont {Riotto}},\ }\href
  {\doibase 10.1103/PhysRevD.78.083517} {\bibfield  {journal} {\bibinfo
  {journal} {Phys. Rev. D}\ }\textbf {\bibinfo {volume} {78}},\ \bibinfo
  {pages} {083517} (\bibinfo {year} {2008})},\ \Eprint
  {http://arxiv.org/abs/0805.3234} {arXiv:0805.3234 [astro-ph]} \BibitemShut
  {NoStop}%
\bibitem [{\citenamefont {Saga}\ \emph {et~al.}(2015)\citenamefont {Saga},
  \citenamefont {Ichiki},\ and\ \citenamefont {Sugiyama}}]{Saga:2014jca}%
  \BibitemOpen
  \bibfield  {author} {\bibinfo {author} {\bibfnamefont {S.}~\bibnamefont
  {Saga}}, \bibinfo {author} {\bibfnamefont {K.}~\bibnamefont {Ichiki}}, \ and\
  \bibinfo {author} {\bibfnamefont {N.}~\bibnamefont {Sugiyama}},\ }\href
  {\doibase 10.1103/PhysRevD.91.024030} {\bibfield  {journal} {\bibinfo
  {journal} {Phys. Rev. D}\ }\textbf {\bibinfo {volume} {91}},\ \bibinfo
  {pages} {024030} (\bibinfo {year} {2015})},\ \Eprint
  {http://arxiv.org/abs/1412.1081} {arXiv:1412.1081 [astro-ph.CO]} \BibitemShut
  {NoStop}%
\bibitem [{\citenamefont {Papanikolaou}\ \emph
  {et~al.}(2021{\natexlab{a}})\citenamefont {Papanikolaou}, \citenamefont
  {Vennin},\ and\ \citenamefont {Langlois}}]{Papanikolaou:2020qtd}%
  \BibitemOpen
  \bibfield  {author} {\bibinfo {author} {\bibfnamefont {T.}~\bibnamefont
  {Papanikolaou}}, \bibinfo {author} {\bibfnamefont {V.}~\bibnamefont
  {Vennin}}, \ and\ \bibinfo {author} {\bibfnamefont {D.}~\bibnamefont
  {Langlois}},\ }\href {\doibase 10.1088/1475-7516/2021/03/053} {\bibfield
  {journal} {\bibinfo  {journal} {JCAP}\ }\textbf {\bibinfo {volume} {03}},\
  \bibinfo {pages} {053} (\bibinfo {year} {2021}{\natexlab{a}})},\ \Eprint
  {http://arxiv.org/abs/2010.11573} {arXiv:2010.11573 [astro-ph.CO]}
  \BibitemShut {NoStop}%
\bibitem [{\citenamefont {Dom\`enech}\ \emph {et~al.}(2020)\citenamefont
  {Dom\`enech}, \citenamefont {Pi},\ and\ \citenamefont
  {Sasaki}}]{Domenech:2020kqm}%
  \BibitemOpen
  \bibfield  {author} {\bibinfo {author} {\bibfnamefont {G.}~\bibnamefont
  {Dom\`enech}}, \bibinfo {author} {\bibfnamefont {S.}~\bibnamefont {Pi}}, \
  and\ \bibinfo {author} {\bibfnamefont {M.}~\bibnamefont {Sasaki}},\ }\href
  {\doibase 10.1088/1475-7516/2020/08/017} {\bibfield  {journal} {\bibinfo
  {journal} {JCAP}\ }\textbf {\bibinfo {volume} {08}},\ \bibinfo {pages} {017}
  (\bibinfo {year} {2020})},\ \Eprint {http://arxiv.org/abs/2005.12314}
  {arXiv:2005.12314 [gr-qc]} \BibitemShut {NoStop}%
\bibitem [{\citenamefont {Dom\`enech}(2020)}]{Domenech:2019quo}%
  \BibitemOpen
  \bibfield  {author} {\bibinfo {author} {\bibfnamefont {G.}~\bibnamefont
  {Dom\`enech}},\ }\href {\doibase 10.1142/S0218271820500285} {\bibfield
  {journal} {\bibinfo  {journal} {Int. J. Mod. Phys. D}\ }\textbf {\bibinfo
  {volume} {29}},\ \bibinfo {pages} {2050028} (\bibinfo {year} {2020})},\
  \Eprint {http://arxiv.org/abs/1912.05583} {arXiv:1912.05583 [gr-qc]}
  \BibitemShut {NoStop}%
\bibitem [{\citenamefont {Inomata}\ \emph
  {et~al.}(2019{\natexlab{a}})\citenamefont {Inomata}, \citenamefont {Kohri},
  \citenamefont {Nakama},\ and\ \citenamefont {Terada}}]{Inomata:2019zqy}%
  \BibitemOpen
  \bibfield  {author} {\bibinfo {author} {\bibfnamefont {K.}~\bibnamefont
  {Inomata}}, \bibinfo {author} {\bibfnamefont {K.}~\bibnamefont {Kohri}},
  \bibinfo {author} {\bibfnamefont {T.}~\bibnamefont {Nakama}}, \ and\ \bibinfo
  {author} {\bibfnamefont {T.}~\bibnamefont {Terada}},\ }\href {\doibase
  10.1088/1475-7516/2019/10/071} {\bibfield  {journal} {\bibinfo  {journal}
  {JCAP}\ }\textbf {\bibinfo {volume} {10}},\ \bibinfo {pages} {071} (\bibinfo
  {year} {2019}{\natexlab{a}})},\ \Eprint {http://arxiv.org/abs/1904.12878}
  {arXiv:1904.12878 [astro-ph.CO]} \BibitemShut {NoStop}%
\bibitem [{\citenamefont {Inomata}\ \emph
  {et~al.}(2019{\natexlab{b}})\citenamefont {Inomata}, \citenamefont {Kohri},
  \citenamefont {Nakama},\ and\ \citenamefont {Terada}}]{Inomata:2019ivs}%
  \BibitemOpen
  \bibfield  {author} {\bibinfo {author} {\bibfnamefont {K.}~\bibnamefont
  {Inomata}}, \bibinfo {author} {\bibfnamefont {K.}~\bibnamefont {Kohri}},
  \bibinfo {author} {\bibfnamefont {T.}~\bibnamefont {Nakama}}, \ and\ \bibinfo
  {author} {\bibfnamefont {T.}~\bibnamefont {Terada}},\ }\href {\doibase
  10.1103/PhysRevD.100.043532} {\bibfield  {journal} {\bibinfo  {journal}
  {Phys. Rev. D}\ }\textbf {\bibinfo {volume} {100}},\ \bibinfo {pages}
  {043532} (\bibinfo {year} {2019}{\natexlab{b}})},\ \Eprint
  {http://arxiv.org/abs/1904.12879} {arXiv:1904.12879 [astro-ph.CO]}
  \BibitemShut {NoStop}%
\bibitem [{\citenamefont {Assadullahi}\ and\ \citenamefont
  {Wands}(2009)}]{Assadullahi:2009nf}%
  \BibitemOpen
  \bibfield  {author} {\bibinfo {author} {\bibfnamefont {H.}~\bibnamefont
  {Assadullahi}}\ and\ \bibinfo {author} {\bibfnamefont {D.}~\bibnamefont
  {Wands}},\ }\href {\doibase 10.1103/PhysRevD.79.083511} {\bibfield  {journal}
  {\bibinfo  {journal} {Phys. Rev. D}\ }\textbf {\bibinfo {volume} {79}},\
  \bibinfo {pages} {083511} (\bibinfo {year} {2009})},\ \Eprint
  {http://arxiv.org/abs/0901.0989} {arXiv:0901.0989 [astro-ph.CO]} \BibitemShut
  {NoStop}%
\bibitem [{\citenamefont {Witkowski}\ \emph {et~al.}(2022)\citenamefont
  {Witkowski}, \citenamefont {Dom\`enech}, \citenamefont {Fumagalli},\ and\
  \citenamefont {Renaux-Petel}}]{Witkowski:2021raz}%
  \BibitemOpen
  \bibfield  {author} {\bibinfo {author} {\bibfnamefont {L.~T.}\ \bibnamefont
  {Witkowski}}, \bibinfo {author} {\bibfnamefont {G.}~\bibnamefont
  {Dom\`enech}}, \bibinfo {author} {\bibfnamefont {J.}~\bibnamefont
  {Fumagalli}}, \ and\ \bibinfo {author} {\bibfnamefont {S.}~\bibnamefont
  {Renaux-Petel}},\ }\href {\doibase 10.1088/1475-7516/2022/05/028} {\bibfield
  {journal} {\bibinfo  {journal} {JCAP}\ }\textbf {\bibinfo {volume} {05}},\
  \bibinfo {pages} {028} (\bibinfo {year} {2022})},\ \Eprint
  {http://arxiv.org/abs/2110.09480} {arXiv:2110.09480 [astro-ph.CO]}
  \BibitemShut {NoStop}%
\bibitem [{\citenamefont {Dalianis}\ and\ \citenamefont
  {Kouvaris}(2021)}]{Dalianis:2020gup}%
  \BibitemOpen
  \bibfield  {author} {\bibinfo {author} {\bibfnamefont {I.}~\bibnamefont
  {Dalianis}}\ and\ \bibinfo {author} {\bibfnamefont {C.}~\bibnamefont
  {Kouvaris}},\ }\href {\doibase 10.1088/1475-7516/2021/07/046} {\bibfield
  {journal} {\bibinfo  {journal} {JCAP}\ }\textbf {\bibinfo {volume} {07}},\
  \bibinfo {pages} {046} (\bibinfo {year} {2021})},\ \Eprint
  {http://arxiv.org/abs/2012.09255} {arXiv:2012.09255 [astro-ph.CO]}
  \BibitemShut {NoStop}%
\bibitem [{\citenamefont {Hajkarim}\ and\ \citenamefont
  {Schaffner-Bielich}(2020)}]{Hajkarim:2019nbx}%
  \BibitemOpen
  \bibfield  {author} {\bibinfo {author} {\bibfnamefont {F.}~\bibnamefont
  {Hajkarim}}\ and\ \bibinfo {author} {\bibfnamefont {J.}~\bibnamefont
  {Schaffner-Bielich}},\ }\href {\doibase 10.1103/PhysRevD.101.043522}
  {\bibfield  {journal} {\bibinfo  {journal} {Phys. Rev. D}\ }\textbf {\bibinfo
  {volume} {101}},\ \bibinfo {pages} {043522} (\bibinfo {year} {2020})},\
  \Eprint {http://arxiv.org/abs/1910.12357} {arXiv:1910.12357 [hep-ph]}
  \BibitemShut {NoStop}%
\bibitem [{\citenamefont {Bernal}\ and\ \citenamefont
  {Hajkarim}(2019)}]{Bernal:2019lpc}%
  \BibitemOpen
  \bibfield  {author} {\bibinfo {author} {\bibfnamefont {N.}~\bibnamefont
  {Bernal}}\ and\ \bibinfo {author} {\bibfnamefont {F.}~\bibnamefont
  {Hajkarim}},\ }\href {\doibase 10.1103/PhysRevD.100.063502} {\bibfield
  {journal} {\bibinfo  {journal} {Phys. Rev. D}\ }\textbf {\bibinfo {volume}
  {100}},\ \bibinfo {pages} {063502} (\bibinfo {year} {2019})},\ \Eprint
  {http://arxiv.org/abs/1905.10410} {arXiv:1905.10410 [astro-ph.CO]}
  \BibitemShut {NoStop}%
\bibitem [{\citenamefont {Das}\ \emph {et~al.}(2022)\citenamefont {Das},
  \citenamefont {Maharana},\ and\ \citenamefont {Muia}}]{Das:2021wad}%
  \BibitemOpen
  \bibfield  {author} {\bibinfo {author} {\bibfnamefont {S.}~\bibnamefont
  {Das}}, \bibinfo {author} {\bibfnamefont {A.}~\bibnamefont {Maharana}}, \
  and\ \bibinfo {author} {\bibfnamefont {F.}~\bibnamefont {Muia}},\ }\href
  {\doibase 10.1093/mnras/stac1620} {\bibfield  {journal} {\bibinfo  {journal}
  {Mon. Not. Roy. Astron. Soc.}\ }\textbf {\bibinfo {volume} {515}},\ \bibinfo
  {pages} {13} (\bibinfo {year} {2022})},\ \Eprint
  {http://arxiv.org/abs/2112.11486} {arXiv:2112.11486 [astro-ph.CO]}
  \BibitemShut {NoStop}%
\bibitem [{\citenamefont {Haque}\ \emph {et~al.}(2021)\citenamefont {Haque},
  \citenamefont {Maity}, \citenamefont {Paul},\ and\ \citenamefont
  {Sriramkumar}}]{Haque:2021dha}%
  \BibitemOpen
  \bibfield  {author} {\bibinfo {author} {\bibfnamefont {M.~R.}\ \bibnamefont
  {Haque}}, \bibinfo {author} {\bibfnamefont {D.}~\bibnamefont {Maity}},
  \bibinfo {author} {\bibfnamefont {T.}~\bibnamefont {Paul}}, \ and\ \bibinfo
  {author} {\bibfnamefont {L.}~\bibnamefont {Sriramkumar}},\ }\href {\doibase
  10.1103/PhysRevD.104.063513} {\bibfield  {journal} {\bibinfo  {journal}
  {Phys. Rev. D}\ }\textbf {\bibinfo {volume} {104}},\ \bibinfo {pages}
  {063513} (\bibinfo {year} {2021})},\ \Eprint
  {http://arxiv.org/abs/2105.09242} {arXiv:2105.09242 [astro-ph.CO]}
  \BibitemShut {NoStop}%
\bibitem [{\citenamefont {Ananda}\ \emph {et~al.}(2008)\citenamefont {Ananda},
  \citenamefont {Carloni},\ and\ \citenamefont {Dunsby}}]{Ananda:2007xh}%
  \BibitemOpen
  \bibfield  {author} {\bibinfo {author} {\bibfnamefont {K.~N.}\ \bibnamefont
  {Ananda}}, \bibinfo {author} {\bibfnamefont {S.}~\bibnamefont {Carloni}}, \
  and\ \bibinfo {author} {\bibfnamefont {P.~K.~S.}\ \bibnamefont {Dunsby}},\
  }\href {\doibase 10.1103/PhysRevD.77.024033} {\bibfield  {journal} {\bibinfo
  {journal} {Phys. Rev. D}\ }\textbf {\bibinfo {volume} {77}},\ \bibinfo
  {pages} {024033} (\bibinfo {year} {2008})},\ \Eprint
  {http://arxiv.org/abs/0708.2258} {arXiv:0708.2258 [gr-qc]} \BibitemShut
  {NoStop}%
\bibitem [{\citenamefont {Papanikolaou}\ \emph
  {et~al.}(2021{\natexlab{b}})\citenamefont {Papanikolaou}, \citenamefont
  {Tzerefos}, \citenamefont {Basilakos},\ and\ \citenamefont
  {Saridakis}}]{Papanikolaou:2021uhe}%
  \BibitemOpen
  \bibfield  {author} {\bibinfo {author} {\bibfnamefont {T.}~\bibnamefont
  {Papanikolaou}}, \bibinfo {author} {\bibfnamefont {C.}~\bibnamefont
  {Tzerefos}}, \bibinfo {author} {\bibfnamefont {S.}~\bibnamefont {Basilakos}},
  \ and\ \bibinfo {author} {\bibfnamefont {E.~N.}\ \bibnamefont {Saridakis}},\
  }\href@noop {} {\  (\bibinfo {year} {2021}{\natexlab{b}})},\ \Eprint
  {http://arxiv.org/abs/2112.15059} {arXiv:2112.15059 [astro-ph.CO]}
  \BibitemShut {NoStop}%
\bibitem [{\citenamefont {Papanikolaou}\ \emph {et~al.}(2022)\citenamefont
  {Papanikolaou}, \citenamefont {Tzerefos}, \citenamefont {Basilakos},\ and\
  \citenamefont {Saridakis}}]{Papanikolaou:2022hkg}%
  \BibitemOpen
  \bibfield  {author} {\bibinfo {author} {\bibfnamefont {T.}~\bibnamefont
  {Papanikolaou}}, \bibinfo {author} {\bibfnamefont {C.}~\bibnamefont
  {Tzerefos}}, \bibinfo {author} {\bibfnamefont {S.}~\bibnamefont {Basilakos}},
  \ and\ \bibinfo {author} {\bibfnamefont {E.~N.}\ \bibnamefont {Saridakis}},\
  }\href@noop {} {\  (\bibinfo {year} {2022})},\ \Eprint
  {http://arxiv.org/abs/2205.06094} {arXiv:2205.06094 [gr-qc]} \BibitemShut
  {NoStop}%
\bibitem [{\citenamefont {Yuan}\ \emph {et~al.}(2019)\citenamefont {Yuan},
  \citenamefont {Chen},\ and\ \citenamefont {Huang}}]{Yuan:2019udt}%
  \BibitemOpen
  \bibfield  {author} {\bibinfo {author} {\bibfnamefont {C.}~\bibnamefont
  {Yuan}}, \bibinfo {author} {\bibfnamefont {Z.-C.}\ \bibnamefont {Chen}}, \
  and\ \bibinfo {author} {\bibfnamefont {Q.-G.}\ \bibnamefont {Huang}},\ }\href
  {\doibase 10.1103/PhysRevD.100.081301} {\bibfield  {journal} {\bibinfo
  {journal} {Phys. Rev. D}\ }\textbf {\bibinfo {volume} {100}},\ \bibinfo
  {pages} {081301} (\bibinfo {year} {2019})},\ \Eprint
  {http://arxiv.org/abs/1906.11549} {arXiv:1906.11549 [astro-ph.CO]}
  \BibitemShut {NoStop}%
\bibitem [{\citenamefont {Zhou}\ \emph {et~al.}(2022)\citenamefont {Zhou},
  \citenamefont {Zhang}, \citenamefont {Zhu},\ and\ \citenamefont
  {Chang}}]{Zhou:2021vcw}%
  \BibitemOpen
  \bibfield  {author} {\bibinfo {author} {\bibfnamefont {J.-Z.}\ \bibnamefont
  {Zhou}}, \bibinfo {author} {\bibfnamefont {X.}~\bibnamefont {Zhang}},
  \bibinfo {author} {\bibfnamefont {Q.-H.}\ \bibnamefont {Zhu}}, \ and\
  \bibinfo {author} {\bibfnamefont {Z.}~\bibnamefont {Chang}},\ }\href
  {\doibase 10.1088/1475-7516/2022/05/013} {\bibfield  {journal} {\bibinfo
  {journal} {JCAP}\ }\textbf {\bibinfo {volume} {05}},\ \bibinfo {pages} {013}
  (\bibinfo {year} {2022})},\ \Eprint {http://arxiv.org/abs/2106.01641}
  {arXiv:2106.01641 [astro-ph.CO]} \BibitemShut {NoStop}%
\bibitem [{\citenamefont {Pitrou}\ \emph {et~al.}(2013)\citenamefont {Pitrou},
  \citenamefont {Roy},\ and\ \citenamefont {Umeh}}]{Pitrou:2013hga}%
  \BibitemOpen
  \bibfield  {author} {\bibinfo {author} {\bibfnamefont {C.}~\bibnamefont
  {Pitrou}}, \bibinfo {author} {\bibfnamefont {X.}~\bibnamefont {Roy}}, \ and\
  \bibinfo {author} {\bibfnamefont {O.}~\bibnamefont {Umeh}},\ }\href {\doibase
  10.1088/0264-9381/30/16/165002} {\bibfield  {journal} {\bibinfo  {journal}
  {Class. Quant. Grav.}\ }\textbf {\bibinfo {volume} {30}},\ \bibinfo {pages}
  {165002} (\bibinfo {year} {2013})},\ \Eprint {http://arxiv.org/abs/1302.6174}
  {arXiv:1302.6174 [astro-ph.CO]} \BibitemShut {NoStop}%
\bibitem [{\citenamefont {Espinosa}\ \emph
  {et~al.}(2018{\natexlab{a}})\citenamefont {Espinosa}, \citenamefont {Racco},\
  and\ \citenamefont {Riotto}}]{Espinosa:2017sgp}%
  \BibitemOpen
  \bibfield  {author} {\bibinfo {author} {\bibfnamefont {J.~R.}\ \bibnamefont
  {Espinosa}}, \bibinfo {author} {\bibfnamefont {D.}~\bibnamefont {Racco}}, \
  and\ \bibinfo {author} {\bibfnamefont {A.}~\bibnamefont {Riotto}},\ }\href
  {\doibase 10.1103/PhysRevLett.120.121301} {\bibfield  {journal} {\bibinfo
  {journal} {Phys. Rev. Lett.}\ }\textbf {\bibinfo {volume} {120}},\ \bibinfo
  {pages} {121301} (\bibinfo {year} {2018}{\natexlab{a}})},\ \Eprint
  {http://arxiv.org/abs/1710.11196} {arXiv:1710.11196 [hep-ph]} \BibitemShut
  {NoStop}%
\bibitem [{\citenamefont {Espinosa}\ \emph
  {et~al.}(2018{\natexlab{b}})\citenamefont {Espinosa}, \citenamefont {Racco},\
  and\ \citenamefont {Riotto}}]{Espinosa:2018eve}%
  \BibitemOpen
  \bibfield  {author} {\bibinfo {author} {\bibfnamefont {J.~R.}\ \bibnamefont
  {Espinosa}}, \bibinfo {author} {\bibfnamefont {D.}~\bibnamefont {Racco}}, \
  and\ \bibinfo {author} {\bibfnamefont {A.}~\bibnamefont {Riotto}},\ }\href
  {\doibase 10.1088/1475-7516/2018/09/012} {\bibfield  {journal} {\bibinfo
  {journal} {JCAP}\ }\textbf {\bibinfo {volume} {09}},\ \bibinfo {pages} {012}
  (\bibinfo {year} {2018}{\natexlab{b}})},\ \Eprint
  {http://arxiv.org/abs/1804.07732} {arXiv:1804.07732 [hep-ph]} \BibitemShut
  {NoStop}%
\bibitem [{\citenamefont {Harada}\ \emph {et~al.}(2013)\citenamefont {Harada},
  \citenamefont {Yoo},\ and\ \citenamefont {Kohri}}]{Harada:2013epa}%
  \BibitemOpen
  \bibfield  {author} {\bibinfo {author} {\bibfnamefont {T.}~\bibnamefont
  {Harada}}, \bibinfo {author} {\bibfnamefont {C.-M.}\ \bibnamefont {Yoo}}, \
  and\ \bibinfo {author} {\bibfnamefont {K.}~\bibnamefont {Kohri}},\ }\href
  {\doibase 10.1103/PhysRevD.88.084051} {\bibfield  {journal} {\bibinfo
  {journal} {Phys. Rev. D}\ }\textbf {\bibinfo {volume} {88}},\ \bibinfo
  {pages} {084051} (\bibinfo {year} {2013})},\ \bibinfo {note} {[Erratum:
  Phys.Rev.D 89, 029903 (2014)]},\ \Eprint {http://arxiv.org/abs/1309.4201}
  {arXiv:1309.4201 [astro-ph.CO]} \BibitemShut {NoStop}%
\bibitem [{\citenamefont {Musco}\ \emph {et~al.}(2009)\citenamefont {Musco},
  \citenamefont {Miller},\ and\ \citenamefont {Polnarev}}]{Musco:2008hv}%
  \BibitemOpen
  \bibfield  {author} {\bibinfo {author} {\bibfnamefont {I.}~\bibnamefont
  {Musco}}, \bibinfo {author} {\bibfnamefont {J.~C.}\ \bibnamefont {Miller}}, \
  and\ \bibinfo {author} {\bibfnamefont {A.~G.}\ \bibnamefont {Polnarev}},\
  }\href {\doibase 10.1088/0264-9381/26/23/235001} {\bibfield  {journal}
  {\bibinfo  {journal} {Class. Quant. Grav.}\ }\textbf {\bibinfo {volume}
  {26}},\ \bibinfo {pages} {235001} (\bibinfo {year} {2009})},\ \Eprint
  {http://arxiv.org/abs/0811.1452} {arXiv:0811.1452 [gr-qc]} \BibitemShut
  {NoStop}%
\bibitem [{\citenamefont {Musco}\ and\ \citenamefont
  {Miller}(2013)}]{Musco:2012au}%
  \BibitemOpen
  \bibfield  {author} {\bibinfo {author} {\bibfnamefont {I.}~\bibnamefont
  {Musco}}\ and\ \bibinfo {author} {\bibfnamefont {J.~C.}\ \bibnamefont
  {Miller}},\ }\href {\doibase 10.1088/0264-9381/30/14/145009} {\bibfield
  {journal} {\bibinfo  {journal} {Class. Quant. Grav.}\ }\textbf {\bibinfo
  {volume} {30}},\ \bibinfo {pages} {145009} (\bibinfo {year} {2013})},\
  \Eprint {http://arxiv.org/abs/1201.2379} {arXiv:1201.2379 [gr-qc]}
  \BibitemShut {NoStop}%
\bibitem [{\citenamefont {Thrane}\ and\ \citenamefont
  {Romano}(2013)}]{Thrane:2013oya}%
  \BibitemOpen
  \bibfield  {author} {\bibinfo {author} {\bibfnamefont {E.}~\bibnamefont
  {Thrane}}\ and\ \bibinfo {author} {\bibfnamefont {J.~D.}\ \bibnamefont
  {Romano}},\ }\href {\doibase 10.1103/PhysRevD.88.124032} {\bibfield
  {journal} {\bibinfo  {journal} {Phys. Rev. D}\ }\textbf {\bibinfo {volume}
  {88}},\ \bibinfo {pages} {124032} (\bibinfo {year} {2013})},\ \Eprint
  {http://arxiv.org/abs/1310.5300} {arXiv:1310.5300 [astro-ph.IM]} \BibitemShut
  {NoStop}%
\bibitem [{\citenamefont {Robson}\ \emph {et~al.}(2019)\citenamefont {Robson},
  \citenamefont {Cornish},\ and\ \citenamefont {Liu}}]{Robson:2018ifk}%
  \BibitemOpen
  \bibfield  {author} {\bibinfo {author} {\bibfnamefont {T.}~\bibnamefont
  {Robson}}, \bibinfo {author} {\bibfnamefont {N.~J.}\ \bibnamefont {Cornish}},
  \ and\ \bibinfo {author} {\bibfnamefont {C.}~\bibnamefont {Liu}},\ }\href
  {\doibase 10.1088/1361-6382/ab1101} {\bibfield  {journal} {\bibinfo
  {journal} {Class. Quant. Grav.}\ }\textbf {\bibinfo {volume} {36}},\ \bibinfo
  {pages} {105011} (\bibinfo {year} {2019})},\ \Eprint
  {http://arxiv.org/abs/1803.01944} {arXiv:1803.01944 [astro-ph.HE]}
  \BibitemShut {NoStop}%
\bibitem [{\citenamefont {Kuroda}\ \emph {et~al.}(2015)\citenamefont {Kuroda},
  \citenamefont {Ni},\ and\ \citenamefont {Pan}}]{Kuroda:2015owv}%
  \BibitemOpen
  \bibfield  {author} {\bibinfo {author} {\bibfnamefont {K.}~\bibnamefont
  {Kuroda}}, \bibinfo {author} {\bibfnamefont {W.-T.}\ \bibnamefont {Ni}}, \
  and\ \bibinfo {author} {\bibfnamefont {W.-P.}\ \bibnamefont {Pan}},\ }\href
  {\doibase 10.1142/S0218271815300311} {\bibfield  {journal} {\bibinfo
  {journal} {Int. J. Mod. Phys. D}\ }\textbf {\bibinfo {volume} {24}},\
  \bibinfo {pages} {1530031} (\bibinfo {year} {2015})},\ \Eprint
  {http://arxiv.org/abs/1511.00231} {arXiv:1511.00231 [gr-qc]} \BibitemShut
  {NoStop}%
\bibitem [{\citenamefont {Siemens}\ \emph {et~al.}(2013)\citenamefont
  {Siemens}, \citenamefont {Ellis}, \citenamefont {Jenet},\ and\ \citenamefont
  {Romano}}]{Siemens:2013zla}%
  \BibitemOpen
  \bibfield  {author} {\bibinfo {author} {\bibfnamefont {X.}~\bibnamefont
  {Siemens}}, \bibinfo {author} {\bibfnamefont {J.}~\bibnamefont {Ellis}},
  \bibinfo {author} {\bibfnamefont {F.}~\bibnamefont {Jenet}}, \ and\ \bibinfo
  {author} {\bibfnamefont {J.~D.}\ \bibnamefont {Romano}},\ }\href {\doibase
  10.1088/0264-9381/30/22/224015} {\bibfield  {journal} {\bibinfo  {journal}
  {Class. Quant. Grav.}\ }\textbf {\bibinfo {volume} {30}},\ \bibinfo {pages}
  {224015} (\bibinfo {year} {2013})},\ \Eprint {http://arxiv.org/abs/1305.3196}
  {arXiv:1305.3196 [astro-ph.IM]} \BibitemShut {NoStop}%
\bibitem [{\citenamefont {Zhao}\ and\ \citenamefont
  {Wang}(2022)}]{Zhao:2022kvz}%
  \BibitemOpen
  \bibfield  {author} {\bibinfo {author} {\bibfnamefont {Z.-C.}\ \bibnamefont
  {Zhao}}\ and\ \bibinfo {author} {\bibfnamefont {S.}~\bibnamefont {Wang}},\
  }\href@noop {} {\  (\bibinfo {year} {2022})},\ \Eprint
  {http://arxiv.org/abs/2211.09450} {arXiv:2211.09450 [astro-ph.CO]}
  \BibitemShut {NoStop}%
\end{thebibliography}%

\begin{widetext}
\newpage

\appendix 
\section{Appendix A: Expressions of the third order transfer functions}\label{sec:A}
\noindent In this appendix, we briefly summarize the explicit expressions of $f_{i}^{(3)}\left(u,\bar{u},\bar{v},\bar{x})\right)$$(i=1,2,3,4)$ \cite{Zhou:2021vcw} \\

\noindent
{\ss \be
f_1^{(3)}(u,\bar{u},\bar{v},x,y)&=&\frac{8}{27} \left(12T_{\phi}(ux) T_{\phi}(\bar{u}y) T_{\phi}(\bar{v}y)-4ux\frac{d}{d(ux)}T_{\phi}(ux) T_{\phi}(\bar{u}y) T_{\phi}(\bar{v}y) \right. \nonumber\\
&-&\frac{2u^2x^2}{3}T_{\phi}(ux) T_{\phi}(\bar{u}y) T_{\phi}(\bar{v}y)-6\bar{u}uxy\frac{d}{d(ux)}T_{\phi}(ux) \frac{d}{d(\bar{u}y)}T_{\phi}(\bar{u}y) T_{\phi}(\bar{v}y)\nonumber\\
&-&\frac{4u^2\bar{u}x^2y}{3}T_{\phi}(ux)\frac{d}{d(\bar{u}y)} T_{\phi}(\bar{u}y) T_{\phi}(\bar{v}y)-4\bar{u}\bar{v}y^2T_{\phi}(ux) \frac{d}{d(\bar{u}y)}T_{\phi}(\bar{u}y) \frac{d}{d(\bar{v}y)}T_{\phi}(\bar{v}y)\nonumber\\
&-&2\bar{u}\bar{v}uy^2x\frac{d}{d(ux)}T_{\phi}(ux) \frac{d}{d(\bar{u}y)}T_{\phi}(\bar{u}y) \frac{d}{d(\bar{v}y)}T_{\phi}(\bar{v}y)\nonumber\\
&-&\left.\frac{2u^2\bar{u}\bar{v}x^2y^2}{3}T_{\phi}(ux) \frac{d}{d(\bar{u}y)}T_{\phi}(\bar{u}y) \frac{d}{d(\bar{v}y)}T_{\phi}(\bar{v}y)\right) \ , \\
	f_2^{(3)}(u,\bar{u},\bar{v},x,y)&=&\frac{8}{27} \left(-6T_{\phi}(ux) T_{\phi}(\bar{u}y) T_{\phi}(\bar{v}y)-4\bar{u}yT_{\phi}(ux) \frac{d}{d(\bar{u}y)}T_{\phi}(\bar{u}y) T_{\phi}(\bar{v}y) \right. \nonumber\\
&-&T_{\phi}(ux)p^2I_{h}^{(2)}(\bar{u},\bar{v},y)-2\bar{u}\bar{v}y^2T_{\phi}(ux) \frac{d}{d(\bar{u}y)}T_{\phi}(\bar{u}y) \frac{d}{d(\bar{v}y)}T_{\phi}(\bar{v}y)\nonumber\\
&+&\frac{u}{v^2x}\frac{d}{d(ux)}T_{\phi}(ux)p^2I_{h}^{(2)}(\bar{u},\bar{v},y)-\frac{u^2}{3v^2} T_{\phi}(ux)p^2I_{h}^{(2)}(\bar{u},\bar{v},y)\nonumber\\
&-&\left.\frac{1-u^2-v^2}{2v^2} T_{\phi}(ux)p^2 I_{h}^{(2)}(\bar{u},\bar{v},y)\right) \ , \\
f_3^{(3)}(u,\bar{u},\bar{v},x,y)&=&\frac{8}{27} \left(\frac{u}{v}\frac{d}{d(ux)}T_{\phi}(ux)pI_V^{(2)}(\bar{u},\bar{v},y)+\frac{y}{8}T_{\phi}(ux)pI_V^{(2)}(\bar{u},\bar{v},y)\right.\nonumber\\
&+&\frac{xuy}{8}\frac{d}{d(ux)}T_{\phi}(ux)pI_V^{(2)}(\bar{u},\bar{v},y)-4T_{\phi}(ux) T_{\phi}(\bar{u}y) T_{\phi}(\bar{v}y)\nonumber\\
&+&8\bar{u}y T_{\phi}(ux) \frac{d}{d(\bar{u}y)}T_{\phi}(\bar{u}y) T_{\phi}(\bar{v}y)+16T_{\phi}(ux) T_{\phi}(\bar{u}y) T_{\phi}(\bar{v}y)\nonumber\\
&+&\left.4\bar{u}\bar{v}y^2T_{\phi}(ux) \frac{d}{d(\bar{u}y)}T_{\phi}(\bar{u}y) \frac{d}{d(\bar{v}y)}T_{\phi}(\bar{v}y)\right) \ , \\
f_4^{(3)}(u,\bar{u},\bar{v},\eta)&=&\frac{8}{27} \left(y\left(T_{\phi}(ux)\frac{\partial}{\partial y}I^{(2)}_{\psi}(\bar{u},\bar{v},y)\right)+uxy\left(\frac{d}{d(ux)}T_{\phi}(ux)\frac{\partial}{\partial y}I^{(2)}_{\psi}(\bar{u},\bar{v},y)\right)\right.\nonumber\\
&+&ux\left(\frac{d}{d(ux)}T_{\phi}(ux)(I^{(2)}_{\psi}(\bar{u},\bar{v},y)+f^{(2)}_{\phi}(\bar{u},\bar{v},y))\right)\nonumber\\
&+&\left.3\left(T_{\phi}(ux)(I^{(2)}_{\psi}(\bar{u},\bar{v},y)+f^{(2)}_{\phi}(\bar{u},\bar{v},y))\right)\right) \ .
\ee}

\section{Appendix B: Detils of the power spectra of third order \acp{SIGW}}\label{sec:B}
\noindent
{\ss \be
	X &=& \left(- 1 + u^2 + v^2 - 2 \bar{u}^2 v^2 +v^2   \bar{v}^2 + u^2 v^2 \bar{v}^2 - v^4\bar{v}^2 + w^2 - u^2 w^2 + v^2 w^2\right)\nonumber\\ 
	&\times& \left(1 - 2 u^2 + u^4 -   2 v^2 - 2 u^2 v^2 + v^4) (1 - 2 v^2 \bar{v}^2 + v^4 \bar{v}^4 - 2 w^2 - 2 v^2  \bar{v}^2 w^2 + w^4 \right)^{-\frac{1}{2}}~, \\ \nonumber\\ 
Y &=& (1 - 2 u^2 + u^4 - 2 v^2 - 2 u^2 v^2 + v^4) (1	- 2 v^2 \bar{v}^2 + v^4 \bar{v}^4 - 2 w^2 - 2	v^2 \bar{v}^2 w^2 + w^4)~,	\\ \nonumber\\
			w_{\pm} & =& \Bigg(\frac{1}{2} + \frac{3}{2 \tilde{k}^2} - \frac{1}{2}
			v^2 \pm \frac{1}{2 v} \sqrt{\left( v^2 -
				\frac{4}{\tilde{k}^2} \right) \left( v^2 - \left( 1 -
				\frac{1}{\tilde{k}} \right)^2 \right) \left( v^2 - \left( 1 +
				\frac{1}{\tilde{k}} \right)^2 \right)}~~\Bigg)^{\frac{1}{2}} \ .
	\ee}
\end{widetext}

\end{document}